\documentclass{jfm}

\usepackage{graphicx}
\usepackage{dcolumn}
\usepackage{bm}

\usepackage{epstopdf}
\graphicspath{{figures/}}
\usepackage[crop=pdfcrop]{pstool}

\usepackage{upgreek}
\usepackage{amsmath}
\usepackage{amsfonts}

\usepackage{siunitx}

\usepackage[latin1]{inputenc}

\usepackage{enumitem}

\usepackage{color}

\newcommand{\xx}{\boldsymbol x}

\newcommand{\vv}{\boldsymbol v}

\begin{document}

\title{Stochastic modelling of a noise driven global instability in a turbulent swirling jet
}











\author
 {
  Moritz Sieber\aff{1} \aff{2}
  \corresp{\email{moritz.sieber@fd.tu-berlin.de}},
	C. Oliver Paschereit \aff{1}
  \and 
  Kilian Oberleithner \aff{2}
  }

\affiliation{\aff{1}Chair of Fluid Dynamics, Institut f$\ddot{\rm{u}}$r Str$\ddot{\rm{o}}$mungsmechanik und Technische
Akustik, \\ Technische Universit$\ddot{\rm{a}}$t Berlin, Berlin,  Germany
\aff{2}Laboratory for Flow Instabilities and Dynamics, Institut f$\ddot{\rm{u}}$r Str$\ddot{\rm{o}}$mungsmechanik und Technische
Akustik, Technische Universit$\ddot{\rm{a}}$t Berlin, Berlin,  Germany
}

\date{\today}

\maketitle

\begin{abstract}

A method is developed to estimate the properties of a global hydrodynamic instability in turbulent flows from measurement data of the limit-cycle oscillations. For this purpose, the flow dynamics are separated in deterministic contributions representing the global mode and a stochastic contribution representing the intrinsic turbulent forcing. Stochastic models are developed that account for the interaction between the two and that allow determining the dynamic properties of the flow from stationary data. The deterministic contributions are modelled by an amplitude equation, which describes the oscillatory dynamics of the instability, and in a second approach by a mean-field model, which additionally captures the interaction between the instability and the mean-flow corrections. The stochastic contributions are considered as coloured noise forcing, representing the spectral characteristics of the stochastic turbulent perturbations. The methodology is applied to a turbulent swirling jet with a dominant global mode. PIV measurements are conducted to ensure that the mode is the most dominant coherent structure and further pressure measurements provide long time series for the model calibration. The supercritical Hopf bifurcation is identified from the linear growthrate of the global mode and the excellent agreement between measured and estimated statistics suggest that the model captures the relevant dynamics. This work demonstrates that the sole observation of limit-cycle oscillations is not sufficient to determine the stability of turbulent flows since the stochastic perturbations obscure the actual bifurcation point. However, the proposed separation of deterministic and stochastic contributions in the dynamical model allows identifying the flow sate from stationary measurements.
\end{abstract}


\section{\label{sec:introduction}Introduction}



\subsection{\label{sec:introduction_intro} General research approach}


Considering a turbulent flow that is dominated by a strong coherent oscillatory motion, the dynamics observed from measurements are twofold. 
There are the deterministic oscillatory motions potentially stemming from an intrinsic global hydrodynamic instability  and the broad-band stochastic motion of the background turbulence.
The differentiation of these deterministic and stochastic dynamics is key for accurately interpreting  and modelling the dominant flow dynamics. 
While this separation can be achieved from a Fourier decomposition, a POD or phase averaging of time-resolved data \citep{Holmes.2012}, it does not reveal the origin of the oscillatory motion independently from the stochastic turbulent forcing. 
Analogously, if oscillations in turbulent flows are modelled by approaches formerly used to describe instabilities in laminar flows, an appropriate closure for the neglected turbulent fluctuations must be included.
In this spirit, the dynamical models developed here built on models that were derived to describe dominant instabilities of laminar flows in the vicinity of the bifurcation point \citep{Stuart.1958}.
However, these models are extended to account for perturbation from background turbulence by adding a stochastic forcing term with the goal of determining the global stability of turbulent flows based only on observational data.

\subsection{\label{sec:introduction_hydr_inst} Linear  instabilities in turbulent flows}

To predict the bifurcation of laminar flows, linear stability analysis has been applied successfully in many cases \citep{Landau.1987}.
The consideration of small perturbations on the base flow and their effect on the eigenmode response of the linearised Navier-Stokes equations provides the decisive exponential growth rate of the coherent structures.
Even for unstable flows, where the instability has already grown considerably, the mode shapes and frequencies of the coherent structures can still be derived from a stability analysis based on the mean-flow field \citep{Barkley.2006}.
This mean-flow stability analysis furthermore allows to assess the sensitivity of the coherent structures to perturbations or forcing of the velocity field \citep{Meliga.2012,Carini.2017}.
In the case of turbulent flows, the coherent fluctuations can be interpreted as linear perturbations of the mean flow, while the remaining fluctuations act as an increased viscosity.
The accurate modelling of the turbulent viscosity is essential to predict the observed coherent structures in highly turbulent flows \citep{Oberleithner.2014,Viola.2014,Tammisola.2016,Rukes.2016c}.

The mean-flow stability analysis has a general trait resulting from the fact that the mean flow is also formed by the Reynolds stresses induced by the investigated coherent structures.
Namely, the predicted amplification rates indicate only neutral stability, no matter how large the original linear instability was that caused the coherent structures to grow.
The investigation of \citet{ManticLugo.2015} demonstrates this property from a coupled simulation using mean-field stability analysis and the steady Navier-Stokes equation to describe the transition of the cylinder wake flow from an unstable stationary state to saturated limit-cycle oscillations of the K\'arm\'an vortex street.
There, it is shown that the Reynolds stresses of the coherent structures change the mean flow such that the instability becomes neutrally stable when the flow reaches the limit-cycle.

The concept of this nonlinearity describing the saturation mechanism of hydrodynamic instabilities is referred to as mean-field theory.
The idea of a weakly nonlinear saturation was first given by Landau's amplitude equation \citep{Landau.1944} derived from analytical reasoning.
In the context of nonlinear stability theory, the mean-field theory explains the observation of supercritical and subcritical Hopf bifurcations via hydrodynamic instabilities \citep{Stuart.1958}.
The work of \cite{Noack.2003} further shows that also the transient dynamics from a steady state to the limit-cycle can be covered by a simple mean-field model including an oscillatory mode for the dynamics of the instability and a shift-mode capturing the slow mean-flow corrections. 
In this work, the model structure and naming are adopted and extended by the consideration of stochastic disturbances induced by background turbulence.

An open aspect of describing transient dynamics by the amplitude equation and the mean-field model is the time-delay between a change of the oscillation amplitude and the resulting correction of the mean flow that leads to a change of the amplification rate. 
\citet{Stuart.1958} assumed this to be an instant feedback which justifies the use of only the amplitude equation without considering the shift-mode. 
However, experimental investigations showed that there are flows that exhibit a delay in this feedback, which motivates the delay-saturation model suggested by \cite{Villermaux.1994}.
In the present work, this delay-saturation model cannot be used because the involved delay operation constitutes a memory of the system that conflicts with the requirements of the stochastic method.
Alternatively, the mean-field model allows considering this delay trough the  inclusion of the shift-mode as an additional state variable. 
Accordingly, we adjust the mean-field model to the dynamics observed in the flow.


\subsection{Stochastic methods for system identification}\label{intro:stochmod}


The identification of the fundamental properties of a physical system from observation data is essential for retaining physical parameters of the flow from the calibrated models.
From the many aspects of this field, we focus on the output-only calibration of grey-box models \citep{Ljung.2012}.
The term output-only refers to the utilisation of observation data of a system.
This approach is in contrast to input-output data that refers to an active forcing of the system and recording of the corresponding response.
The term grey-box model refers to the use of empirical models that are motivated by physics or the observed dynamics.
This is in contrast to white-box models which are derived directly from the governing equations and also in contrast to black-box models that only reproduce dynamics and are not related to the physics of the system.

The models employed in this work for the system identification are the amplitude equation and the mean-field model. 
Having two and three state variables, respectively, these are low-order models that represent only the dominant dynamics of a turbulent flow. 
The remaining dynamics are considered as stochastic turbulent fluctuations that may enter the model as a stochastic forcing. 
From the experimental perspective, an output-only calibration is performed that uses only stationary measurements of the flow. 
However, the intrinsic forcing of the flow by its background turbulence is also considered for the system identification.
This requires to estimate the properties of the stochastic forcing in line with the deterministic dynamics of the flow.

The accurate treatment of such stochastic equations requires a systematic introduction of the model.
An overview of the topic can be found in the review of \cite{Friedrich.2011} about stochastic methods for data-driven identification of physical systems.
The general concept of the approach is a strict separation of deterministic and stochastic contributions in the model.
This is obtained by requiring the model to have the form of a Langevine equation.
Therefore, the model must account for all the deterministic dynamics contained in the data.
It is not possible to oversimplify the model and lump secondary dynamics to the stochastic forcing.
The stochastic part must be uncorrelated from the deterministic part to handle the data in this framework.

The requirement of the stochastic differential equation having the form of a Langevine equation implies that the future evolution of the system depends only on the current state of the system.
Therefore, there should be no memory of the system or hidden variables that interfere with the resolved state variables.
Moreover, the stochastic component of the equation must be uncorrelated such that the evolution of the system behaves like a Markov process.
All deterministic dynamics present in the observed system must be described by the model since they cannot be lumped together with the unresolved stochastic turbulence.
However, a sufficient separation of time-scales allows treating some of the dynamics as stochastic contributions.
The assumption of a Markov process allows describing the temporal evolution of the probability density function (PDF) by the corresponding Fokker-Planck equation.
This eliminates the stochastic variable and gives a probabilistic description of the system that can be compared to statistical moments obtained from measured data.

The basic principle of stochastic methods proposed by \cite{Friedrich.2000} utilises the direct computation of the drift and diffusion terms from the first and second statistical moments of the data.
This requires to conduct a limiting process that may conflict with the non-vanishing correlation time of the stochastic process \citep{Lehle.2018}.
An alternative is the evaluation of finite time propagation of the PDF with the Fokker-Planck equation and comparison with the PDF of the data \citep{Kleinhans.2007} or the direct estimation of the parameters from the adjoined Fokker-Planck equation \citep{Boujo.2017}.
Furthermore, the stationary PDF of the data can be compared to the stationary solution of the Fokker-Planck equation as pursued by \cite{Noiray.2013}, \cite{Bonciolini.2017} and \cite{Lee.2019}.

Stochastic methods were applied to identify universal features of the turbulent cascade \citep{Friedrich.1997, Reinke.2018}, leading to very simple models that correctly capture the spectral properties of the cascade.
Concerning the thermoacoustic system of a combustor, \cite{Noiray.2013} proposed various approaches to derive the parameters of a Van der Pol oscillator from different measures of the data.
This was further extended to handle also systems with non-white noise \citep{Bonciolini.2017}.
The existence of coherence resonance in a thermoacoustic system was also shown by \cite{Kabiraj.2015}, where the external stochastic forcing allowed to further classify the associated Hopf bifurcation before the onset of the instability \citep{Saurabh.2017}.

The stochastic dynamics of turbulent axisymmetric and bluff-body wakes were studied by \cite{Rigas.2015} and \cite{Brackston.2016}, respectively, showing that the symmetry-breaking modes are governed by simple stochastic models.
The dynamics of a freely rotating disc in uniform flow was shown by \cite{Boujo.2019} to be governed by a stochastic low-order model that captures the main features.
The self-excited oscillations in the fluid acoustic system of bottle whistling were furthermore shown to be governed by a randomly forced Van der Pol oscillator \citep{Boujo.2020}.
The investigations of \cite{Zhu.2019} and \cite{Lee.2019} revealed that the bifurcation, leading to the global instability of a low-density jet, can be characterised from the evaluation of stochastic forcing in the stable regime of the flow. 

Beyond these very recent investigations, there were previous approaches to model the dynamics of turbulent flows by stochastic equations. For example, the description of the \textit{bursts} in boundary layers as noisy heteroclinic cycles by \cite{Stone.1989} and the control of such dynamics by \cite{Coller.1994}. The occurrence of noise-induced dynamics of marginally stable modes is often also referred to as coherence resonance referring to the work of \cite{Gang.1993}.
Another probabilistic description of fluid dynamics was recently proposed by \cite{Kaiser.2014}, who used data-driven state-space segmentation to identify the related transition probability that allows inferring the dynamics of the flow.
In the work of \cite{Brunton.2016}, the residuals of the model are interpreted as intermittent forcing of a deterministic system, this is in contrast to the present approach, where Markov properties of the stochastic forcing are expected.

Besides the work of \cite{Zhu.2019} and \cite{Lee.2019}, there are no applications of stochastic methods to describe the characteristics of a global hydrodynamic instability.
In contrast to their work, the current investigation does not rely on external forcing of the flow but utilises the background turbulence as intrinsic stochastic forcing.


\subsection{Detailed research approach}


In the present work, we consider the dominant coherent structure occurring in turbulent jets at high swirl.
Swirling jets are commonly used in combustors to provide anchoring of the flame \citep{Syred.1974}.
This is due to the unique feature of the flow which is known as vortex breakdown.
If the swirl intensity in the jet exceeds a certain threshold, the jet breaks down and forms a recirculation region in the centre \citep{Billant.1998}.
In combustion applications, this provides recirculation of hot exhausts that stabilises the flame.
The swirl intensity that governs the onset of vortex breakdown is quantified by the swirl number, given by the ratio of azimuthal to axial momentum flux \citep{Chigier.1967} (see also appendix \ref{sec:swirl_number} for the definition).

Beyond the onset of vortex breakdown, the swirl number remains the major control parameter for a global hydrodynamic instability.
With increasing swirl number, the flow passes through a supercritical Hopf bifurcation that gives rise to a global mode \citep{Gallaire.2003,Liang.2005,Oberleithner.2011}.
It takes the form of a single helical structure that precesses around the centre axis of the jet.
In the following, this specific global mode is referred to as a {\itshape helical mode}.
In combustion-related applications, the helical mode is also known as a {\itshape precessing vortex core} \citep{Syred.1974,Terhaar.2014,Vanierschot.2019}.

The supercritical Hopf bifurcation of global modes are commonly described by the Landau equation \citep{Landau.1944}. Accordingly, the limit-cycle amplitude of the helical mode $|A_\mathrm{LC}|$ should be proportional to the swirl number $S$ as
\begin{align}
 |A_\mathrm{LC}|^2 \propto S-S_c,
\end{align}
with the swirl number being the control parameter that governs the instability of the global mode.
The critical swirl number $S_c$ marks the bifurcation point, where the flow transitions from a stable to an unstable state.

\begin{figure}
    \centering
            \includegraphics{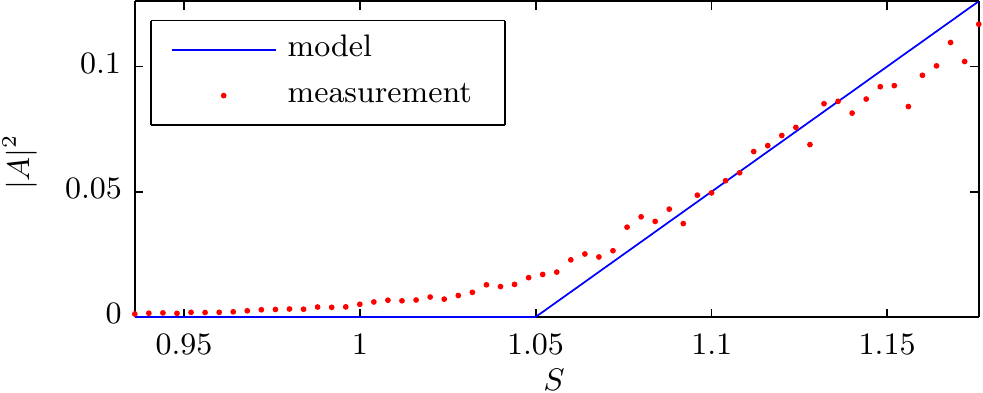}
    \caption{Bifurcation diagram for a supercritical Hopf Bifurcation as described by the Landau equation. The solid line gives the limit-cycle of the model and the dots indicate measurements from a turbulent flow.}
    \label{fig:bifurcation_plots}
\end{figure}

However, in turbulent flows, the helical mode dynamics are subjected to stochastic turbulent forcing, which leads to a deviation from the Landau model in the vicinity of the bifurcation point. This is shown in figure~\ref{fig:bifurcation_plots}, where measurements in a turbulent swirling flow reveal a continuous increase of the helical mode amplitude at potentially subcritical conditions.
This observation calls for stochastic methods to develop a dynamical system that explains the diffusion of the observed dynamics. 
Such a model differentiates between helical mode activity due to an intrinsic flow instability or due to turbulent forcing, providing a clear description of the dynamical flow state that enables the identification of the bifurcation point.

The immediate approach to handle these stochastic contributions is the investigation of the amplitude equation with additive noise, and the estimation of equation parameters by inspection of the stationary probability density function (PDF) of measured amplitudes.
This is also pursued in the present investigations as the first attempt.
Therefore, an analytical expression for the stationary PDF is derived from the amplitude equation and calibrated from measured PDFs as done in the investigations of \cite{Noiray.2013}, \cite{Bonciolini.2017} and \cite{Lee.2019}.

For reasons of simplicity, the noise in the stochastic equation is usually assumed to be white, which is not the case for turbulent perturbations.
This property is addressed in this work by the use of coloured noise created by an Ornstein--Uhlenbeck (OU) process \citep{Hanggi.1994}.
The effect of the noise properties on the accuracy of the system identification is assessed from a numerical model study, as done by \cite{Bonciolini.2017}.
Similar to their study, a sufficient separation of noise and model time-scales is found to be decisive for the reliability of the approach. 

Furthermore, a mean-field model is incorporated in the dynamical system to capture the mean-field corrections related to the saturation of the instability on the limit-cycle. 
This provides a deeper insight into the interaction between deterministic and stochastic dynamics.
However, the system identification is adapted since no analytical expression for the stationary PDF was obtained.
Therefore, it is calibrated and adapted to fit directly the estimated dynamics derived from the data.
This is in contrast to the amplitude equation, which can be solved analytically and compared to well-converged PDFs, allowing an application to a large class of problems.  

The remainder of this work is organised in the following way. In section \ref{sec:methods} we outline the main experimental methods and the identification of coherent structures from the data and the dynamic content of the flow is presented from a decomposition of PIV and pressure measurements. 
In section~\ref{sec:methods2}, the stochastic amplitude equation and the corresponding system identification are described, followed by the validation of the approach based on a numerical study and the calibration of the model from the experimental data.
Section~\ref{sec:mean-field_model} shows the stochastic mean-field model and its calibration from measurement data.
Finally, the main findings are summarised in section \ref{sec:conclusions}.


\section{\label{sec:methods} Experimental details and observations of swirling jet dynamics}



\subsection{\label{sec:methods_exp} Experimental setup and data acquisition}

\begin{figure}
    \centering
            \includegraphics[width=0.80\textwidth]{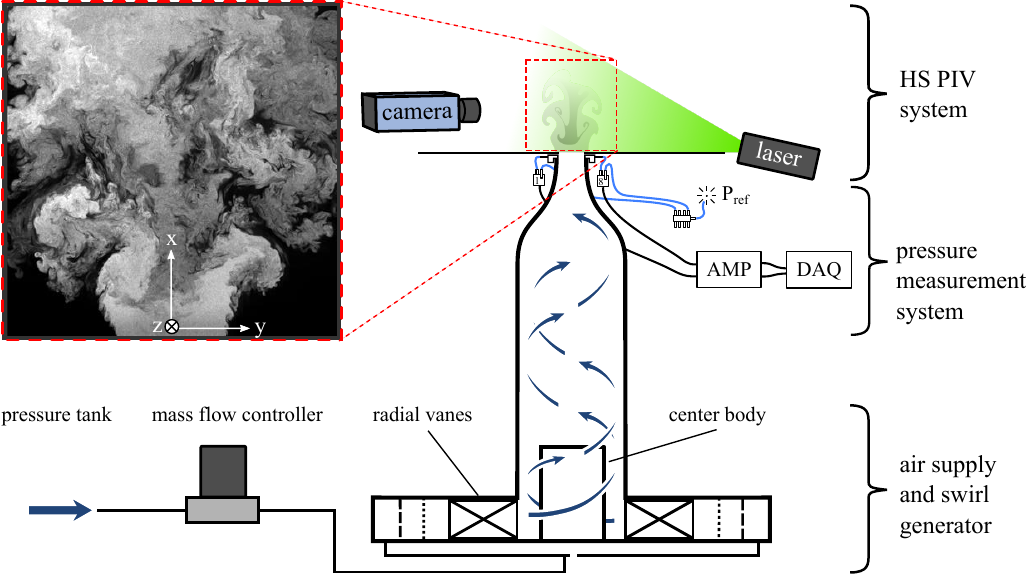}
    \caption{Schematic of the experimental setup and the utilised measurement systems. The magnified picture of the flow field is an experimental dye visualisation of the jet.}
    \label{fig:experimental_setup}
\end{figure}

Experiments are conducted using a generic swirling jet setup, shown schematically  in figure~\ref{fig:experimental_setup} together with the measurement devices used. 
It consists of an adjustable radial swirl generator that is supplied at four azimuthal positions by regulated air from a pressure reservoir. Several grids ensure a homogeneous distribution of the supplied air. 
The swirling air is guided normally to the swirler plane in a circular duct of \SI{150}{\milli\meter} diameter followed by a contoured contraction reducing the diameter to $D=\SI{51}{\milli\meter}$. 
The jet is emanating into a large space (\SI{4}{\meter} wide, \SI{3}{\meter} high) that constitute unconfined boundary conditions in radial and streamwise direction. 
Further details of the experimental setup can be found in  \cite{Rukes.2015,Muller.2020}.

The jet is investigated at a fixed mass flow of \SI{50}{\kg\per\hour} that results in a nozzle bulk velocity of $u_{\mathrm{bulk}} = 5.7\si{\meter\per\second}$ and a respective Reynolds number of \num{20,000} based on the nozzle diameter $D$. 
The swirl intensity is adjusted by an automated stepper motor that controls the angle of the swirler vanes. 
Swirl numbers in the range of \num{0.8} to \num{1.35} were tested. 
The current swirl number definition is based on the linear fit of the integral swirl number against the angle of the swirler vanes. 
The reason for this approach is due to difficulties with a consistent formulation of the integral swirl number across the investigated swirl range, which is further detailed in appendix \ref{sec:swirl_number}.

The velocity field of the swirling jet is captured by high-speed particle image velocimetry (PIV) as indicated in figure \ref{fig:experimental_setup}. 
A meridional section of the jet is illuminated by the laser and recorded by a  high-speed camera. 
The flow is seeded by silicon-oil droplets (DEHS:Bis(2-ethylhexyl) sebacate) of a nominal diameter smaller than \SI{5}{\micro\meter} which are added to the air between mass flow controller and swirler. 
For each configuration, a set of \num{2000} images was recorded at a rate of \SI{1}{\kHz}. 
The images were processed with PIVview (PIV{\textit{TEC}} GmbH) using standard PIV processing \citep{Willert.1991} enhanced by iterative multigrid interrogation \citep{Soria.1996} and image deformation \citep{Huang.1993}.

The high-speed PIV captured only the axial and radial velocity component, which is sufficient for the determination of coherent structures. 
The time resolution of the measurement, however, is essential for the later analysis.
The computation of the swirl number, however, requires also the mean azimuthal velocity component. It was determined from non-time-resolved stereoscopic PIV measurements conducted in a previous investigation \citep{Rukes.2015}.

Together with the PIV acquisition, pressure measurements were conducted. 
The probes are located at eight positions around the circumference of the nozzle lip, which are referred to in the following as $p_k$ $k=1\ldots8$. 
The piezoresistive sensors with a range of \SI{1}{\kilo\pascal} were amplified with an in-house bridge amplifier and recorded with a \num{24}bit AD converter at a rate of  \SI{2}{\kHz}. 
The resonance frequency of the sensor-tubing-system was at \SI{400}{Hz}, which is acceptable for the conducted investigations, where the oscillations of the dominant mode were in the range of \SI{50}{Hz} to \SI{80}{Hz}.
The resonance of the sensor caused an amplification of the signal by \SI{2}{\percent} at \SI{50}{Hz} and \SI{4}{\percent} at \SI{80}{Hz}.
Other than the PIV, the pressure was recorded for \SI{60}{\sec} to achieve converged statistics.
Pressure measurements were conducted in an automated procedure, where the swirler angel was increased in steps of \SI{0.1}{\degree} corresponding to $\Delta S$ of \num{4e-3}.


\subsection{\label{sec:methods_identification} Identification of coherent structures from measurement data}


This section details how the helical mode is identified from the PIV and pressure data.
The first part covers the extraction of the dominant coherent structures from the PIV data using the spectral proper orthogonal decomposition (SPOD) as described by \citet{Sieber.2016b}.
The method allows for a modal decomposition of the velocity  snapshots, reading  
\begin{equation}
\vv(\xx,t) = \overline{\vv}(\xx) + \vv'(\xx,t) = \overline{\vv}(\xx) + \sum_{i=1}^{N}{a_{i}(t)\Phi_{i}(\xx)},
\end{equation}
where the fluctuating part of the velocity field $\vv'$ is expressed as a sum of spatial modes $\Phi$ and time coefficients $a$.
The used variant of the SPOD provides a time-domain representation of the decomposition, which is essential for the present analysis of flow dynamics. 
This is in contrast to a related frequency domain representation  \citep{Towne.2018}, which is not applicable here.

The SPOD has promising potential for the analysis of time-domain phenomena in turbulent flows \citep{Noack.2016b}.
This has been shown for the identification of dynamics in the flow of a separated airfoil \citep{Ribeiro.2017}, the transient interaction and switching between different dynamics in a combustor \citep{Stohr.2017}, or the provision of proper dynamics for the autonomous modelling of flow dynamics \citep{Lui.2019}. 
The principal advantage of the SPOD in these applications is that the dynamics are separated according to their space-time coherence while the time information is maintained, allowing the time-domain analysis of the individual dynamics.

The SPOD is based on the snapshot POD proposed by \citet{Sirovich.1987}, with the extensions that the snapshot correlation matrix
\begin{align}
R_{i,j} = \frac{1}{N}\left<\vv'(\xx,t_i),\vv'(\xx,t_j)\right>
\end{align}
is filtered, which constrains the spectral bandwidth of the coefficients.
The advantage of this approach against spectral filtering of data before or after performing the POD is that the filtering is implicitly handled by the decomposition.
Hence, there is no need to specify central frequencies of individual phenomena within the data or to apply digital filters.
Instead, only the filter width $N_f$ must be chosen, as it sets the bandwidth of the coefficients.
The filter is implemented as a simple convolution filter of the snapshot correlation matrix
\begin{align}
S_{i,j} = \sum_{k=-N_f}^{N_f}{g_k R_{i+k,j+k}}. \label{eqn:SPOD_filter}
\end{align}
where $g_k$ is a Gaussian filter kernel.
Other than snapshot POD, the SPOD needs time-resolved data to allow time-domain filtering of the data.
Further details on the selection of the filter size and handling of boundary conditions can be found in \cite{Sieber.2016b, Sieber.2017}.
Beyond the application of the filter to the correlation matrix, the procedure is the same as for the snapshot POD.
In the present investigation, a filter width $N_f$ of two times the oscillation period is used.

For further analysis, the link of mode pairs in the decomposition is identified from a cross-spectral correlation among all possible pairs of coefficients (harmonic correlation as in \cite{Sieber.2016b}).
The appearance of coherent structures as pairs is commonly observed in the POD of real-valued input data.
It can be understood as the real and imaginary part of a mode that is obtained from a spectral analysis.
In the following, mode pairs (e.g. SPOD mode 1 and 2) are treated as one mode and they are coupled for the time domain analysis as the real and imaginary part of a complex coefficient $A(t) = a_1(t) + i a_2(t)$.

In the present application, the pressure measurements are used for the dynamic modelling since they allow much longer time series resulting in better-converged statistics than the PIV snapshots.
The agreement of mode amplitudes determined from PIV and pressure measurements is detailed in appendix \ref{sec:PIV_pressure_relation}.
To obtain the mode amplitude, the pressure measurements from the eight positions around the nozzle lip are decomposed azimuthally into Fourier modes, according to  
\begin{align}
  &\widehat{p}_m(t) = \frac{1}{8}\sum_{k=1}^{8} p_k(t) \mathrm{e}^{-i m k \pi/4 } \label{eqn:fourier_decomp},
\end{align}
where $m$ indicates the azimuthal mode number.
The instantaneous amplitude of the helical mode with azimuthal wavenumber $m=1$ is then given as $A(t) =\widehat{p}_1(t)$, accounting for the single-helical shape of this coherent structure.
The signal was filtered around the average oscillation frequency of the helical mode $f_o$ in the band $[\frac{2}{3} f_o, \frac{3}{2} f_o]$.
To set the centre of the filter band, the frequency of the mode was identified from the peak in the unfiltered spectrum and also compared to the frequency of the SPOD coefficients from the PIV measurements.
In the low swirl regime where no peak was visible, the frequency was extrapolated from lager swirl numbers.

In addition to the oscillatory mode, the dynamics of the slow-varying contributions to the flow are obtained from the pressure measurements. 
They are represented by the shift-mode that is determined from the $m=0$ pressure mode: $B(t) = \widehat{p}_0(t)$.
Thereby, a relation of the mean flow and the mean pressure is assumed which is supported by the data shown in appendices \ref{sec:swirl_number} and \ref{sec:PIV_pressure_relation}.
The $\widehat{p}_0$ signal is low pass filtered at $4f_o$ to remove acoustic perturbations from the upstream duct.
All pressure signals are normalised by the maximum amplitude of the helical mode, being \SI{6}{\pascal}, to ease the numerical procedures and improve the readability of graphs.
The normalisation is not related to the dynamic pressure which would be around \SI{20}{\pascal} in the present case.


\subsection{\label{sec:results_PIV} Flow field and dominant coherent structures}

\begin{figure}
    \centering
            \includegraphics[width=1\textwidth]{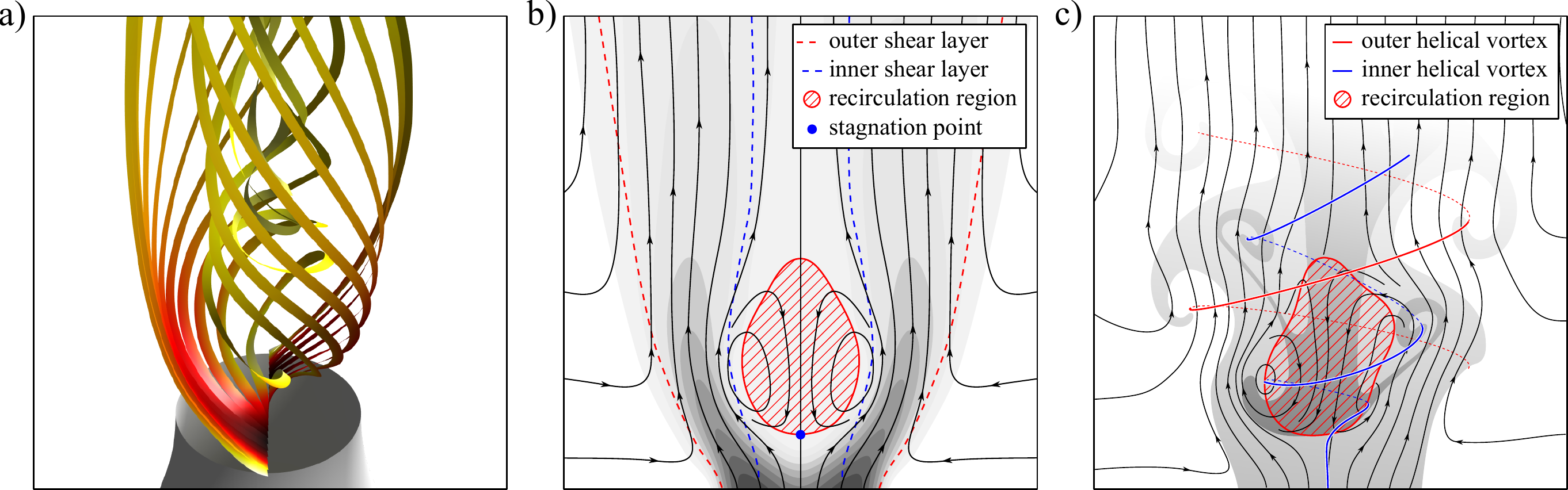}
    \caption{Schematics of the flow field of a swirling jet: a) stream lines of the mean velocity field coloured by velocity magnitude, b) slice through the symmetry axis of the mean velocity field represented by sectional streamlines and velocity magnitude as gray contour levels, c) instantaneous velocity field indicated by sectional streamlines and coherent structure as gray background. Specific features of the flow fields are marked and indicated in the legends. The breakdown bubble is indicated by the recirculating flow in the centre.}
    \label{fig:flow_field}
\end{figure}

The mean flow field and the structure of the global mode in the swirling jet are sketched in figure \ref{fig:flow_field}.
Due to vortex breakdown, the flow forms an annular jet with inner and outer shear layers.
The global mode manifests in the central vortex of the jet that meanders and wraps around the breakdown bubble in a helical shape.
The roll-up of the outer shear layer is synchronised with this motion resulting in a secondary helical vortex.
These counter winding helical vortices cause a spiralling, flapping motion of the annular jet.
With respect to the mean flow, the helical vortices are co-rotating and counter winding.
In the current section, the global mode is referred to as helical mode, in contrast to secondary dynamics that are also observed in the data. 
The helical mode is considered as the oscillatory mode in the amplitude equation.

\begin{figure}
    \centering
     \includegraphics[width=\textwidth]{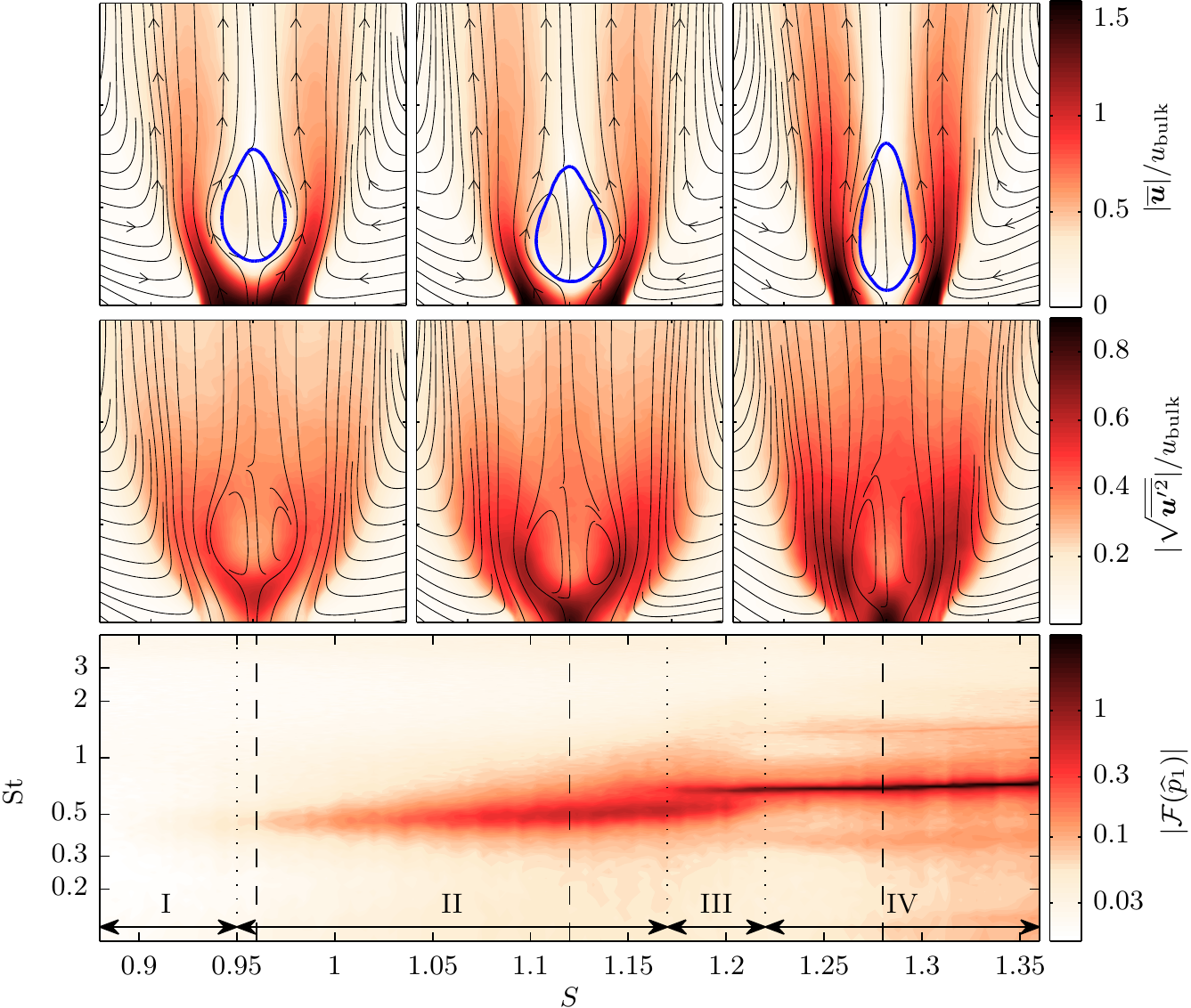}
    \caption{Mean flow and spectrum of the dominant oscillatory mode at different swirl numbers. The top row shows contours of relative velocity magnitude together with streamlines and a thick blue line indicating zero axial velocity. The mid-row shows the relative turbulence intensity as contours together with streamlines. The bottom plot shows the magnitude of the spectrum of the pressure Fourier mode $\widehat{p}_1$ with a log scaled colour-map for all investigated swirl numbers across the relevant Strouhal number range. The dashed vertical lines indicate the respective swirl number ($S=[0.96,1.12,1.28]$) of the mean flow plots above. The arrows and vertical dotted lines indicate different regimes in the swirl number rage.}
    \label{fig:mean_flow_spectrum}
\end{figure}

The velocity field and pressure spectra across the investigated swirl range are collectively presented in figure \ref{fig:mean_flow_spectrum}. 
The velocity is normalised with the bulk velocity and the frequency is normalised with the bulk velocity and nozzle diameter to obtain Stouhal numbers as $\mathrm{St} =\frac{f_o D}{u_{\mathrm{bulk}}}$.

The mean velocity fields (figure \ref{fig:mean_flow_spectrum} top row) show typical features of a swirling jet undergoing vortex breakdown as sketched in figure \ref{fig:flow_field}.
From the low to the medium swirl number case there is only little change of the velocity field. 
The breakdown bubble mainly moves upstream closer to the nozzle in connection with an increased jet spreading.
At large swirl number, the breakdown bubble becomes narrower and longer, while the jet spreading decreases.

The turbulence intensity (figure \ref{fig:mean_flow_spectrum} mid row) constantly increases with the swirl strength.
The peak intensity is concentrated in the shear layers, while the value at the upstream stagnation point is always the largest.
With increasing swirl the shear layers become thicker and the fluctuations become spread over a larger area.

Several different regimes are visible from the spectrum of the helical mode (figure \ref{fig:mean_flow_spectrum} bottom). Starting from low swirl, there is a range where no oscillations are observed. 
At swirl numbers of approximately \num{0.95} there is a slight peak in the spectrum at Strouhal numbers around \num{0.5}. 
With increasing swirl, the intensity of the peak increases as well as the frequency. 
Throughout the range, the spectral peak is rather broad.
At swirl number of \num{1.17}, a second, narrow peak at slightly higher frequency appears which continues to grow while the previous one fades out.
After the first peak disappeared at a swirl number of \num{1.22}, the second peak further grows and the frequency increases slightly.
At the upper end of the range, the Strouhal number is at \num{0.73}.

Based on the characteristics of the pressure spectra, the investigated swirl range can be divided into four regimes as indicated in the bottom plot of figure \ref{fig:mean_flow_spectrum}.
Regime I corresponds to the swirl range where no helical mode dynamics are observable. 
In regime II, the helical mode appears with a broad spectral peak and the mean flow shows a continuous change. Regime III corresponds to a bi-stable condition. It is characterised by intermittent switching between the states in regimes II and IV, which is visible from time-series data not shown here. Regime IV shows a similar trend as regime II, but with a sharper spectral peak and a longer breakdown bubble.

\begin{figure}
    \centering
            \includegraphics{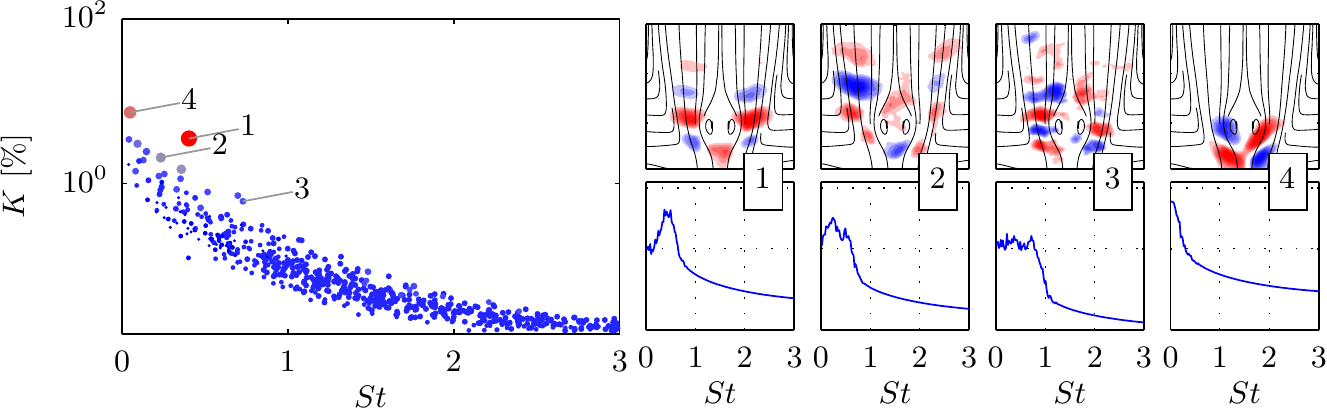}\\
            \includegraphics{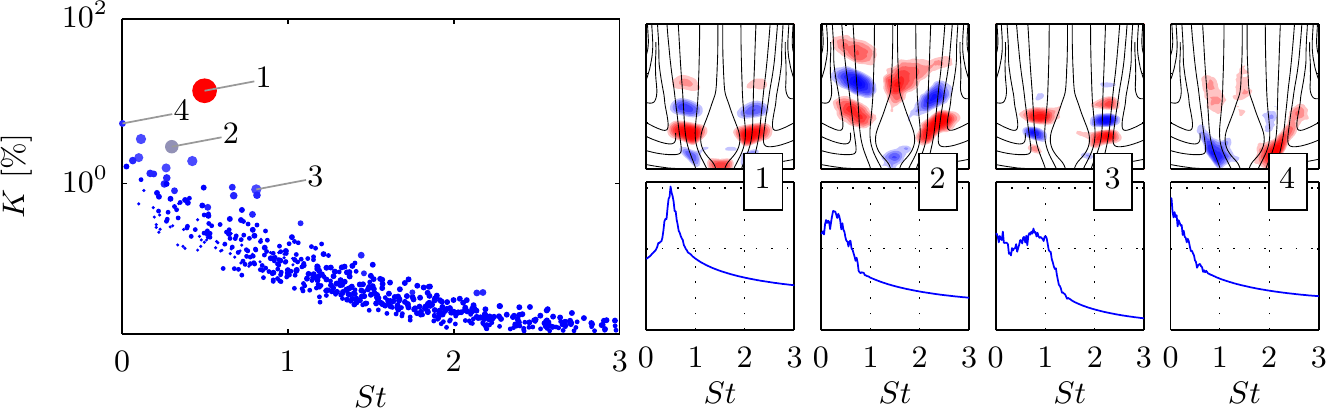}\\
            \includegraphics{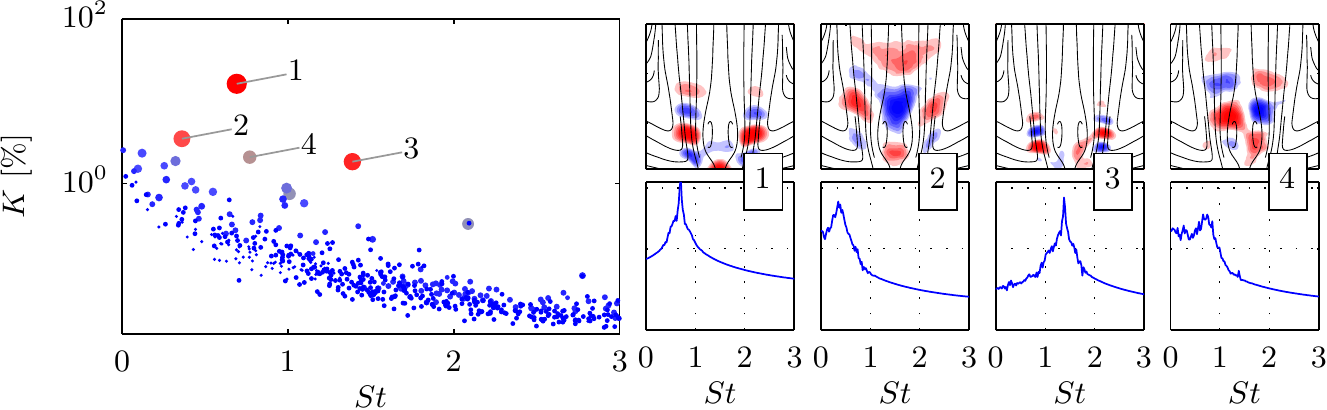}
    \caption{SPOD spectrum (left) and spatial modes with mode coefficient spectrum (right) for $S=[0.96,1.12,1.28]$ (from top to bottom). The spectrum shows mode energy against Strouhal number, where each dot corresponds to a mode pair and the colour/size indicated the spectral coherence. For selected modes the spatial and spectral content is detailed as indicated by the numbers. The spatial modes are shown by the transverse velocity together with streamlines of the mean flow. The mode coefficients are given as power spectral density.}
    \label{fig:SPOD_2}
\end{figure}

To provide an overview of the range of dynamics observed in the flow,  SPOD is conducted based on flow snapshots acquired at three selected swirl numbers. The cases correspond to the mean velocity fields given in figure \ref{fig:mean_flow_spectrum}.
The SPOD spectrum presented  in figure \ref{fig:SPOD_2} shows all modes contained in the flow. 
The detailed picture of spatial structures and coefficient spectra are given for some modes, which are selected such that a consistent ordering is obtained for the different swirl numbers. 

Figure \ref{fig:SPOD_2} (top-row) shows the SPOD results for the low swirl case $S=0.96$, which corresponds to the onset of the helical mode. The first SPOD mode (\#1) shows a clear peak at $St=0.4$ and the corresponding spatial structure shows typical characteristics of the helical mode \citep{Rukes.2015}.
Another dominant structure (mode \#4) with low frequency ($St=0.05$) is also prominent.
It corresponds to an axial movement of the breakdown bubble as also observed previously by \cite{Rukes.2015}. 
The other inspected modes (\#2 and \#3) do not indicate clear structures but might be precursors of the dynamics that become more pronounced at higher swirl numbers.

The medium swirl case $S=1.12$ (figure \ref{fig:SPOD_2}, mid-row) shows the helical mode again as the most dominant structure (mode \#1). 
There is little change in the mode shape compared to the low swirl case, but its relative energy increased significantly and it oscillates at a higher frequency of $St=0.5$.
There exists another prominent coherent structure at $St=0.3$ (mode \#2) that resembles a helical mode in the wake of the breakdown bubble, which has already been observed in combustor flows \citep{Sieber.2016d,Terhaar.2014}.
The location in the wake and the single helical mode structure indicates a strong relation to the global mode that is observed in laminar swirling flows, known as spiral vortex breakdown \citep{Ruith.2003,Gallaire.2006,Qadri.2013}.
In the following, this structure is called the wake mode.
At $St=0.8$ there is another mode that slightly sticks out from the continuous spectrum (mode \#3). 
The spatial structure does not exhibit clear symmetries and the coefficient spectrum shows two peaks at $St=0.8$ and $St=1.0$.
This mode supposedly represents the helical modes second harmonic together with an interaction between the wake mode and the helical mode.
The mode at very low frequency (mode \#4) corresponds to the slow-varying changes of the flow, which is still considerable.

For the high swirl case $S=1.28$ (figure \ref{fig:SPOD_2}, bottom row), the mode dynamic is very clear and the low-frequency modes are reduced.
The helical mode (\#1) at $St=0.69$ has further gained in energy and now clearly exhibits higher harmonics at $St=1.38$ (mode \#3) and $St=2.1$.
The wake mode (\#2) has consolidated at $St=0.36$ and also exhibits a higher harmonic at $St=0.77$ (mode \#4).
At the frequency of $St=1.05$, where interactions between helical mode and wake mode are expected, there is an agglomeration of less dominant modes.
This indicates an interaction between both structures, which, however, is not captured by a single mode.

Overall, the SPOD of the flow shows only little change of the helical mode shape with increasing swirl number.
The observed increase in energy and the shift of the oscillation frequency is consistent with the pressure spectra (figure \ref{fig:mean_flow_spectrum}).
The variations of the mean flow decrease with increasing swirl, which is probably due to the proximity of the breakdown bubble to the nozzle lip at high swirl, which constraints its movement. 
The SPOD reveals other modes such as the wake mode and higher harmonics, but the helical mode is most dominant for the entire swirl range.
The identified interaction modes remain weak with energies of at least one order of magnitude less than the helical mode.
Therefore, the interaction between the helical mode and subdominant modes can be neglected in the following analysis. 

\section{\label{sec:methods2} Modelling and system identification by the amplitude equation}


The dynamics of the helical mode are modelled by a stochastic amplitude equation.
The corresponding model is derived at the beginning of the following section.
This is followed by the consideration of coloured noise forcing and the derivation of the calibration procedure from data.
Thereafter, the reliability of the model is investigated from a numerical study and the model is calibrated from experimental data.


\subsection{\label{sec:methods_model} Model design for the stochastic amplitude equation}


The form of the presented model is based on the stochastic methods described by \cite{Friedrich.2011}. 
Accordingly, the stochastic differential equation that describes the temporal evolution of a system has the form of a Langevine equation
\begin {align}
\dot{X} = f(X) + g(X) \xi, \label{eqn:basic_langevine}
\end{align}
where the dot indicates the time derivative of the state $X$. 
The deterministic contributions are collected in the drift $f$ and stochastic contributions are covered by the diffusion $g$.
The stochastic forcing $\xi$ is Gaussian white noise with vanishing correlation time $\left<\xi(t)\xi(s)\right> = \delta(t-s)$ ($\delta$ being the Dirac delta function).
The vanishing correlation, which was initially referred to as a Markov process, is essential for the later analysis of the equation.

The amplitude equation serves here as the most simple model to describe the dynamics of the helical mode and the leading-order nonlinearity that governs the saturation at the limit-cycle \citep{Stuart.1958, Landau.1987}.
It is given as
\begin{align}
\dot{A} = (\sigma + i \omega) A - \alpha \left|A\right|^2A \label{eq:landau} + ...,
\end{align}
where $A$ is a complex variable that describes the modal coefficient of a periodic perturbation of the flow field. 
The model parameters are the oscillation frequency $\omega$, amplification rate $\sigma$ and the Landau constant $\alpha$. 
Here, no change of frequency due to saturation and no higher-order contributions are considered. 
Higher-order terms are neglected since we expect a supercritical Hopf bifurcation as previously observed by \cite{Oberleithner.2011} rather than a sub-critical bifurcation that would require a higher-order model \citep{Meliga.2012b}.

For the stochastic flow model, noise is added to the amplitude equation \eqref{eq:landau}, reading 
\begin{align}
\dot{A} = (\sigma + i \omega) A - \alpha \left|A\right|^2A + \xi, \label{eq:landau_forced}
\end{align}
where the noise is complex $\xi = \xi_r + i \xi_i$ with uncorrelated real and imaginary part $\left<\xi_{r}\xi_{i}\right>=0$. 
The noise has zero mean and the variance $\Gamma = \overline{\xi \xi^*}$ ($^*$ indicates the complex conjugate). A complex-valued noise is necessary to maintain the symmetry of the equation.

The representation of the mode coefficient as amplitude and phase $A = |A| e^{i\phi}$ allows to separate the unperturbed amplitude equation\eqref{eq:landau} into two simple real valued equations reading
\begin{align}
\dot{|A|} = \sigma|A| - \alpha |A|^3 \ ; \ \ \dot{\phi} = \omega. \label{eq:landau_pol}
\end{align}
The amplitude and phase are regarded as slow and fast variables meaning that $\sigma \ll \omega$.
In combination with the amplitude and phase representation, this allows stochastic averaging over the fast variables of the stochastic equation \eqref{eq:landau_forced} \citep{Roberts.1986} that gives
\begin{align}
&\dot{|A|} = \sigma|A| - \alpha |A|^3  + \frac{\Gamma}{|A|} + \xi_{A} \label{eq:ampl_stoch}\\ 
&\dot{\phi} = \omega + \frac{\xi_{\phi}}{|A|} \label{eq:phase_stoch}.
\end{align}
The conversion results in different contributions of the noise to the amplitude and phase ($\xi_{A}$ and $\xi_{\phi}$) as indicated by the subscripts. These noise contributions are real valued, uncorrelated $\left<\xi_{A}\xi_{\phi}\right>=0$ and have halve the variance of the complex perturbation $\left<\xi_{A},\xi_{A}\right>=\left<\xi_{\phi},\xi_{\phi}\right>=\Gamma/2$.

The stochastic averaging of the equation refers to a short-term integration of the stochastic contributions that is analogue to previous approaches \citep{Noiray.2017, Lee.2019}, where the Van der Pol oscillator instead of the amplitude equation was considered.
The Van der Pol oscillator has an equivalent amplitude and phase representation as \eqref{eq:ampl_stoch}-\eqref{eq:phase_stoch} which makes the approaches comparable. 
The only difference in the present approach is the use of complex noise that maintains the symmetry of the equation even for large amplitudes and strong noise.

The change of variables separates the model into two equations.
The amplitude equation \eqref{eq:ampl_stoch} is independent of the phase, describing the exponential amplification and saturation mechanism of the oscillatory mode.
In the unperturbed case \eqref{eq:landau_pol}, the sign of the amplification rate $\sigma$ determines whether the oscillations occur or not.
However, the additional deterministic contribution $\frac{\Gamma}{|A|}$ in the stochastic amplitude equation \eqref{eq:ampl_stoch} prevents the oscillator from decaying to zero amplitude even for negative amplification rates.
In addition to the new deterministic contribution, the additive stochastic perturbation remains in the amplitude equation.
Considering the phase equation \eqref{eq:phase_stoch}, the stochastic forcing causes the phase to be dependent not only on the frequency but also on parametric perturbations that scale inversely with the the amplitude.
Therefore, the stochastic forcing couples the phase to the amplitude equation.

To estimate the noise intensity from experimental data, the phase equation \eqref{eq:phase_stoch} is rearranged to 
\begin{align}
\xi_{\phi,\mathrm{est}} = |A|\left(\dot{\phi} - \omega\right) \label{eq:phase_distortion},
\end{align}
which can be used to estimate the noise intensity using
\begin{align}
\frac{\Gamma_\mathrm{est}}{2} = \left<\xi_{\phi,\mathrm{est}},\xi_{\phi,\mathrm{est}}\right>\label{eq:noise_ampl_est}.
\end{align}
The extraction of the remaining parameters from the amplitude equation is detailed in the subsequent section.


\subsection{\label{sec:colored_noise} Stochastic perturbations with coloured noise}


The application of stochastic methods requires the noise to be white (uncorrelated) to obtain a Langevine equation. 
However, the noise may have a short correlation time, which makes the stochastic forcing coloured instead of white noise.
This is reasonable as long as the time scales of the deterministic and stochastic dynamics are well separated \citep{Hanggi.1994}.
A simple type of coloured noise is obtained from an Ornstein--Uhlenbeck (OU) process which has the autocorrelation
\begin{align}
\left<\xi(t),\xi(s)\right> = \frac{D}{\tau}e^{-|t-s|/\tau} \label{eq:OU_corr}
\end{align}
and is created from a basic stochastic process as
\begin{align}
\dot{\xi} = - \frac{1}{\tau} \xi + \frac{\sqrt{D}}{\tau} \xi_{w}.\label{eq:OU_process}
\end{align}
In the above equations $\xi$ denotes the noise created from the OU process with  time scale $\tau$ and intensity $D$. 
The variable $\xi_{w}$ denotes a white noise process with a variance of one that drives the OU process.
To incorporate the coloured noise into the stochastic flow model, the specific correlation \eqref{eq:OU_corr} is included in the stochastic averaging of the amplitude equation \citep{Noiray.2017, Bonciolini.2017}.
Accordingly, it is sufficient to replace the noise intensity in \ref{eq:ampl_stoch} by an effective noise intensity
\begin{align}
\Gamma_{\tau} = \frac{2D}{\tau^2\omega^2+1} = \frac{2 \tau \left<\xi,\xi\right>}{\tau^2\omega^2+1},\label{eq:col_noise_intens}
\end{align}
which is equivalent to the power spectrum  of the OU noise at the oscillator frequency $\omega$.
Note that the factor two in the equation results from the two noise components in the complex-valued forcing. 

The time scale and intensity of the stochastic forcing were estimated from the following empirical relation
\begin{align}
\left<\xi_\phi(t),\xi_\phi(s)\right> \approx \frac{D}{2\tau}e^{|t-s|/\tau}\cos(\omega (t-s)). \label{eq:col_noise_ampl_est}
\end{align}
There, the phase distortion according to \eqref{eq:phase_distortion} is compared to the analytical correlation of an OU process \eqref{eq:OU_corr}. 
The additional cosine term accounts for the different coordinates the noise is represented in.
The noise adds to the amplitude equation as uncorrelated real and imaginary parts, whereas the phase distortion is considered in the rotating coordinates of the oscillator.

The procedure outlined here, using an effective noise and the estimation of the noise properties from the phase distortion, is only strictly valid for noise time scales that are smaller than the oscillator time scale and for small noise amplitudes.
This limit of applicability is further investigated in section \ref{sec:results_numeric}.


\subsection{\label{sec:methods_stat_fit} System identification from stationary probability density functions}


The method presented here is used to estimate the parameters of the amplitude equation by inspecting the stationary probability density function (PDF) of measured amplitudes.
An analytical expression for the stationary PDF is derived from the amplitude equation \eqref{eq:ampl_stoch} through the corresponding Fokker-Plank equation.
Therefore, the amplitude equation is brought in the form of the Langevine equation \eqref{eqn:basic_langevine}, where the corresponding drift and diffusion terms are
\begin{align}
f(|A|) &= |A| \left(\sigma - \alpha |A|^2 \right)\  + \frac{\Gamma}{|A|} \\
g(|A|) &= \sqrt{\Gamma/2},
\end{align}
respectively.
The assumption of a Markov process allows to describe the temporal evolution of the PDF of the magnitude $|A|$ by the the Fokker-Planck equation \citep{Friedrich.2011}
\begin {align}
\frac{\partial}{\partial t}P(|A|,t) = -\frac{\partial}{\partial |A|}\left( f(|A|) P(|A|,t) \right) +  \frac{1}{2} \frac{\partial^2}{\partial |A|^2}\left( g^2(|A|) P(|A|,t) \right), \label{eq:FPE}
\end{align}
where $P$ refers to the PDF.
This eliminates the stochastic variable and gives a probabilistic description of the system, which can be compared to statistical moments obtained from measured data.

In order to identify the model parameters, a stationary solution of the Fokker-Planck equation $\frac{\partial}{\partial t}P(|A|,t) = 0$ has to be found.
Following the derivations of \cite{Noiray.2017}, the stationary Fokker-Planck equation with vanishing PDF at infinite amplitudes and constant diffusion $g$ simplifies to
\begin{align}
\frac{\mathrm{d}}{\mathrm{d} |A|}P(|A|) - \frac{2}{g^2} f(|A|) P(|A|) = 0.
\end{align}
The corresponding solution is obtained by writing the drift equation in a potential form, reading
\begin {align}
f(|A|) =  -\frac{\partial\mathcal{V}(|A|)}{\partial |A|}\ \ \mathrm{with} \ \ \mathcal{V}(|A|) = -\frac{\sigma}{2}|A|^2 + \frac{\alpha}{4}|A|^4 - \Gamma\ln{|A|},
\end{align}
which gives
\begin {align}
P(|A|) = \mathcal{N}\exp\left(-\frac{4}{\Gamma}\mathcal{V}(|A|)\right).
\end{align}
The scale parameter $\mathcal{N}$ is chosen to normalise the PDF such that $\int_0^\infty P(|A|)\mathrm{d}|A| = 1$.
The expression contains a mixture of stochastic and deterministic parameters that are identified from different measures.

The strategy for the presented analysis is the following. 
We use a generic  model of the PDF
\begin{align}
P_{mod}(|A|) = \mathcal{N} |A| \exp\left(c_1 |A|^2 + c_2 |A|^4 \right) \label{eqn:amplitude_PDF_model}
\end{align}
with the parameters $c_1$ and $c_2$. 
It is fit to the experimental PDF $P_{exp}$ using the Kulback-Leibler divergence 
\begin{align}
D_{KL} = \int_0^\infty{P_{exp}(|A|)\log\left(\frac{P_{exp}(|A|)}{P_{mod}(|A|)}\right)} \mathrm{d}|A|\label{eqn:KLdiv}
\end{align}
as similarity measure. 
The divergence $D_{KL}$ is minimised using common numerical procedures. 
The effective noise intensity $\Gamma_\tau$ is obtained from the procedure described in section \ref{sec:colored_noise}.
Together with the fit parameters of the model PDF, this provides the physical model parameters
\begin{align}
\sigma = \frac{c_1\Gamma_\tau}{2}  \ \ \mathrm{and} \ \ 
\alpha = -c_2 \Gamma_\tau \label{eq:param_est},
\end{align}
where the estimation of the amplification rate $\sigma$ is most important for the interpretation of the flow state.

\begin{figure}
    \centering
            \includegraphics{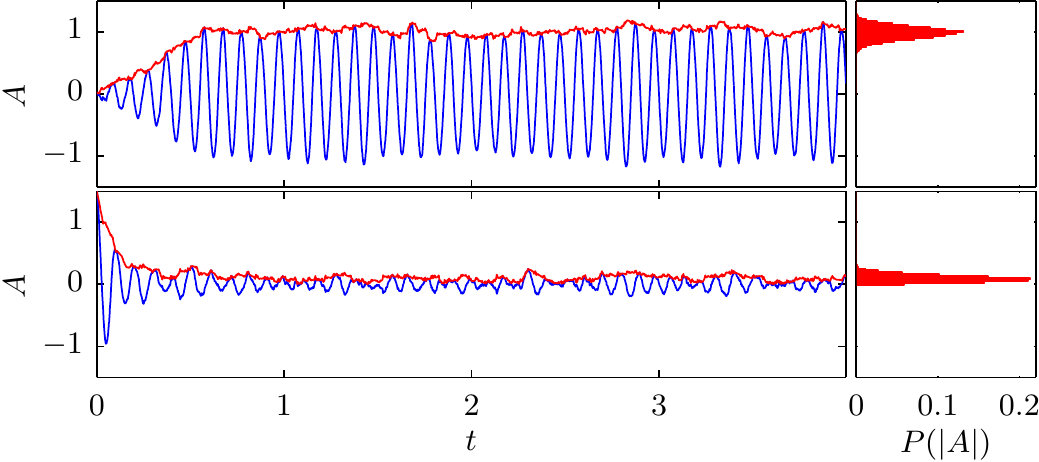}
    \caption{Exemplary time series of the simulated stochastic amplitude equation \eqref{eq:landau_forced} for an unstable system (upper plot) and a stable system (lower plot). The blue lines show the real part of $A$ and the red line gives the corresponding magnitude $|A|$. The bar blots on the right show the stationary probability density function of the magnitude. The models are simulated with white noise and the ratio $\omega/\sigma$ is 10 for the unstable and -10 for the stable case.}
    \label{fig:simulate_system}
\end{figure}

For illustrative purposes, figure \ref{fig:simulate_system} shows a characteristic time series of an unstable system ($\sigma>0$) and a stable system ($\sigma<0$) that are both subjected to stochastic forcing. 
The signal was generated by numeric integration of the forced amplitude equation \eqref{eq:landau_forced}.
The unstable system is initialised at $A=0$ and quickly approaches the limit-cycle.
Due to the stochastic forcing, it never settles at the limit-cycle. 
The stationary PDF of the amplitude has the expected value at the limit-cycle and a Gaussian-like distribution. 
The distribution reflects the balance between the stabilising drift that pushes the system to the limit-cycle and the destabilising stochastic forcing.
This is the same for the stable system, with the difference that the deterministic dynamics of the system tends to zero amplitude but gets continuously excited by the stochastic forcing.
The estimation of the model parameters from the stationary PDF and the phase distortion allows to quantify the balance between these forces and provides estimates of physical amplification rates.

\subsection{\label{sec:results_numeric} Numerical study of the amplitude equation with coloured noise}


The numerical study aims to validate the assumptions made during the derivation of the analytical procedure in sections \ref{sec:methods_model} to \ref{sec:methods_stat_fit}.
Especially the stochastic averaging \eqref{eq:ampl_stoch} and the estimation of the noise properties \eqref{eq:col_noise_ampl_est} need to be validated for coloured noise.
Therefore, the original complex-valued equation \eqref{eq:landau_forced} is numerically integrated in time with coloured noise from an OU process \eqref{eq:OU_process}.
The OU process is integrated with a second-order Runge-Kutta scheme for stochastic equations \citep{Tocino.2002}.
The amplitude equation is integrated with a fourth-order Runge-Kutta scheme with fixed time steps corresponding to \num{100} steps per oscillation period.
The coloured noise is simulated with twice the temporal resolution to allow the Runge-Kutta scheme to be evaluated for intermediate time steps.
In the cases with white noise forcing, the amplitude equation is simulated like the OU process.
The amplitude equation is simulated for a range of amplification rates $\sigma = -1 \ldots \SI{1}{\per\second}$, noise time scales $\tau = 0 \ldots \SI{1}{\second}$ and effective noise intensities $\Gamma_{\tau} = 0.01 \ldots 1$.
The frequency is kept constant at $f_o=\SI{10}{Hz}$ as well as the Landau constant $\alpha =1$.
The parameters of the amplitude equation are identified according to equation \eqref{eq:param_est} and the effective noise intensity according to equation \eqref{eq:col_noise_intens} and \eqref{eq:col_noise_ampl_est}.

\begin{figure}
    \centering
    \includegraphics{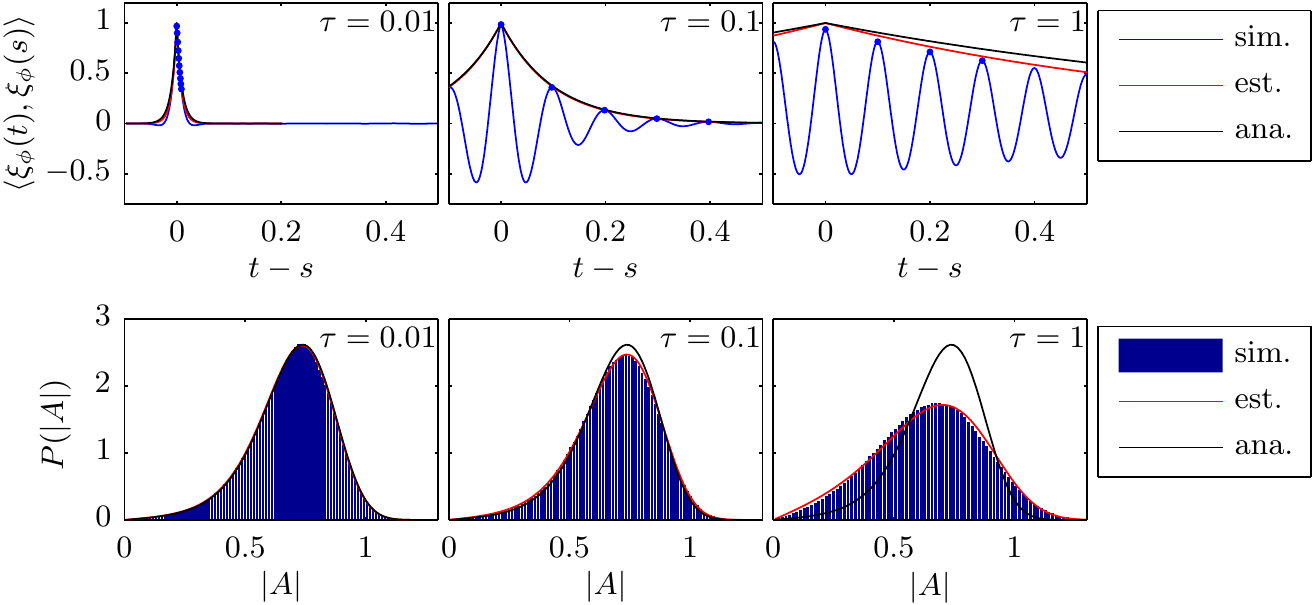}
    \caption{Selected simulation results of the numerical study at $\sigma=0.5$, $\Gamma_{\tau} = 0.1$ and  $\tau = [0.01,0.1,1]$, as indicated in the up-right corner. The top row shows noise correlation from the simulation, as well as the estimated and analytical decay (plots are normalised by the value at zero shift $t=s$). The blue dots indicate the local maxima which are used to estimate the decay. The bottom row shows the  PDF of the envelope calculated from the simulation, the estimated fit to the simulation results and the expected analytical PDF.}
    \label{fig:model_study_fit_detail}
\end{figure}

The fit of the model to the simulation data is exemplified in figure \ref{fig:model_study_fit_detail} for selected cases. 
The selection shows an unstable system for different noise time scales but the same effective noise intensity.
In the top row of figure \ref{fig:model_study_fit_detail} the autocorrelation of the phase distortion is shown. 
Accordingly, the estimation of the noise parameters from the decay of the correlation envelope works well for small and intermediate noise time scales but deviates for the large noise time scale.
The bottom row of figure \ref{fig:model_study_fit_detail} shows the simulated, estimated and analytically derived amplitude PDFs.
Since all presented simulations have the same effective noise intensity and model parameters, the analytic PDF must be the same for all cases (see equation \eqref{eq:col_noise_ampl_est}).
For the small noise time scale ($\tau=0.01$) all shown PDFs agree very well. For the moderate noise time scale ($\tau=0.1$) the analytic and simulated PDFs deviate slightly. 
Note that the noise time scale here is of the same order as the oscillation frequency.
Therefore, the assumed separation of time scales is no longer met, but nevertheless, the deviations stay small.
The long noise time scale ($\tau=1$) causes significant deviations of the expected and simulated PDFs. 
Furthermore, the shape of the PDF is not correctly captured by the analytical model, indicating the need for higher-order approximations.

\begin{figure}
    \centering
    \includegraphics{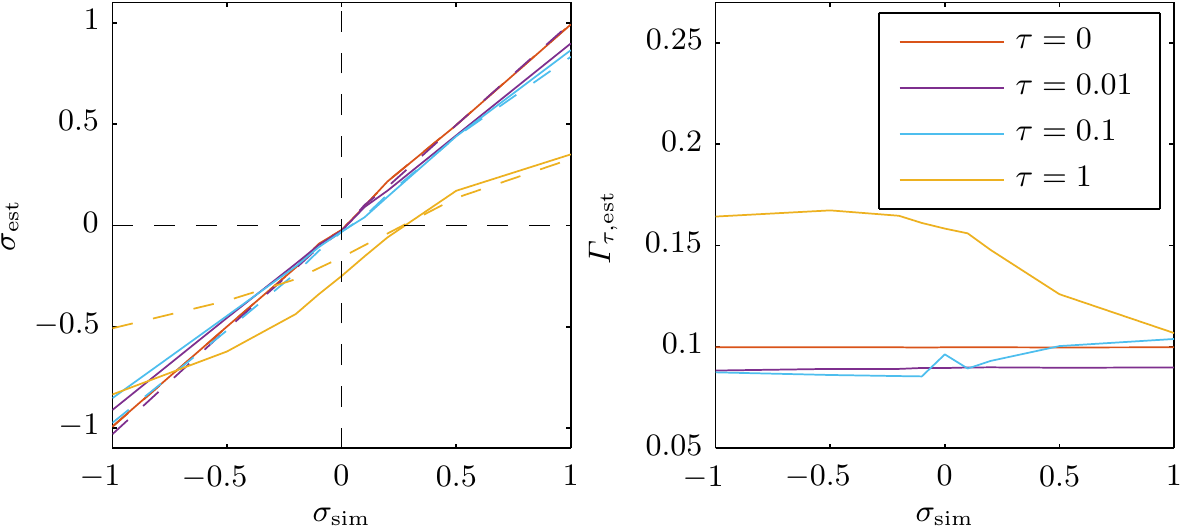}
    \caption{Estimates of the model coefficients from the simulation data at $\Gamma_{\tau} = 0.1$ and different noise time scales indicated in the legend. The left plot shows the estimated against the true amplification rate. The solid curves show results with estimated $\Gamma_{\tau}$ and the dashed lines with true $\Gamma_{\tau} = 0.1$. The plot on the right shows the corresponding estimation of the effective noise intensity from data.}
    \label{fig:model_study_combined}
\end{figure}

The estimated amplification rates and effective noise intensities for all simulated cases are presented in figure \ref{fig:model_study_combined}. 
The estimated amplification rate is displayed against the one used in the simulation. 
For a perfect estimation, the graph displays a straight diagonal, which is the case for the simulation with white noise ($\tau=0$). 
For $\tau = 0.01$ and $\tau = 0.1$ the relative deviations stay below \SI{10}{\percent}, while for $\tau = 1$ the deviations exceed this limit.
The source for the error are either an inaccurate fit of the PDF with the model equation \eqref{eqn:amplitude_PDF_model} or an inaccurate estimation of the noise properties from the empirical correlation function \eqref{eq:col_noise_ampl_est}.
To differentiate between these two errors sources, the estimates are also shown based on the true noise intensity.
Accordingly, the errors from both sources are of the same order of magnitude.
For the cases with $\tau \le 0.1$, the error generally only changes the slope of the amplification curve and the bifurcation point (zero crossing) is accurately captured.
With increasing noise time scales, the absolute rate becomes increasingly underpredicted, as seen from the decreasing slope.
The estimation for $\tau = 1$ strongly deviates for the PDF fit as well as for the noise estimation.
Hence, the model is not able to cover cases with noise time scales larger than the oscillation period $\tau > 1/f_o$.

The numerical study shows that the model and the simplifications made in the deviations are valid as long as the noise time scale stays below the oscillatory time scale. 
Surprisingly, there is no need for a large separation of time scales if small deviations are acceptable. 
A perfect agreement, however, is only possible for purely white noise.


\subsection{\label{sec:results_1D_fit} Parameter estimation of the amplitude equation from experiments}

\begin{figure}
    \centering
    \includegraphics{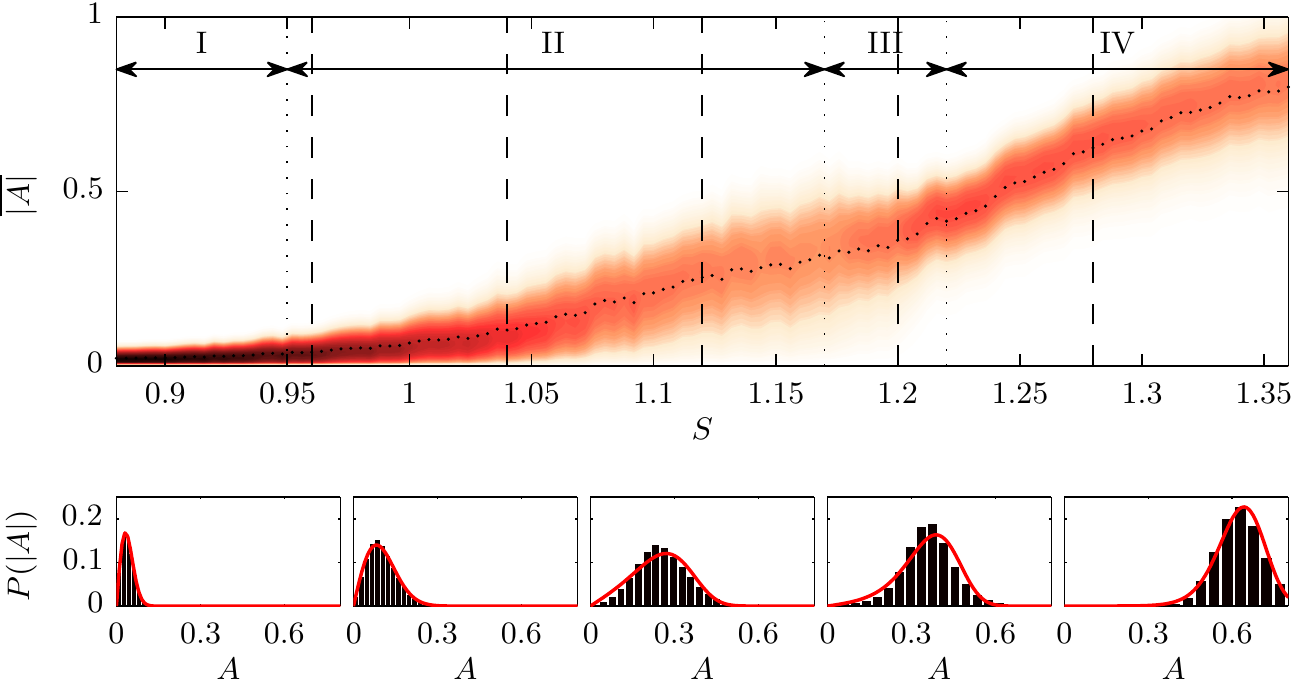}
    \caption{Analysis of the amplitude statistics derived from pressure measurement. The top plot shows the PDF of the envelope $P(|A|)$ as contours for different swirl numbers. The second row shows bar plots of the measurement data together with the best fit of the theoretical PDF as a red line. These plots correspond to sections of the contour plot above, where the corresponding positions are marked as dashed lines. The arrows and vertical dotted lines indicate different regimes in the swirl number rage.}
    \label{fig:pressure_prob_hist}
\end{figure}
\begin{figure}
    \centering
    \includegraphics{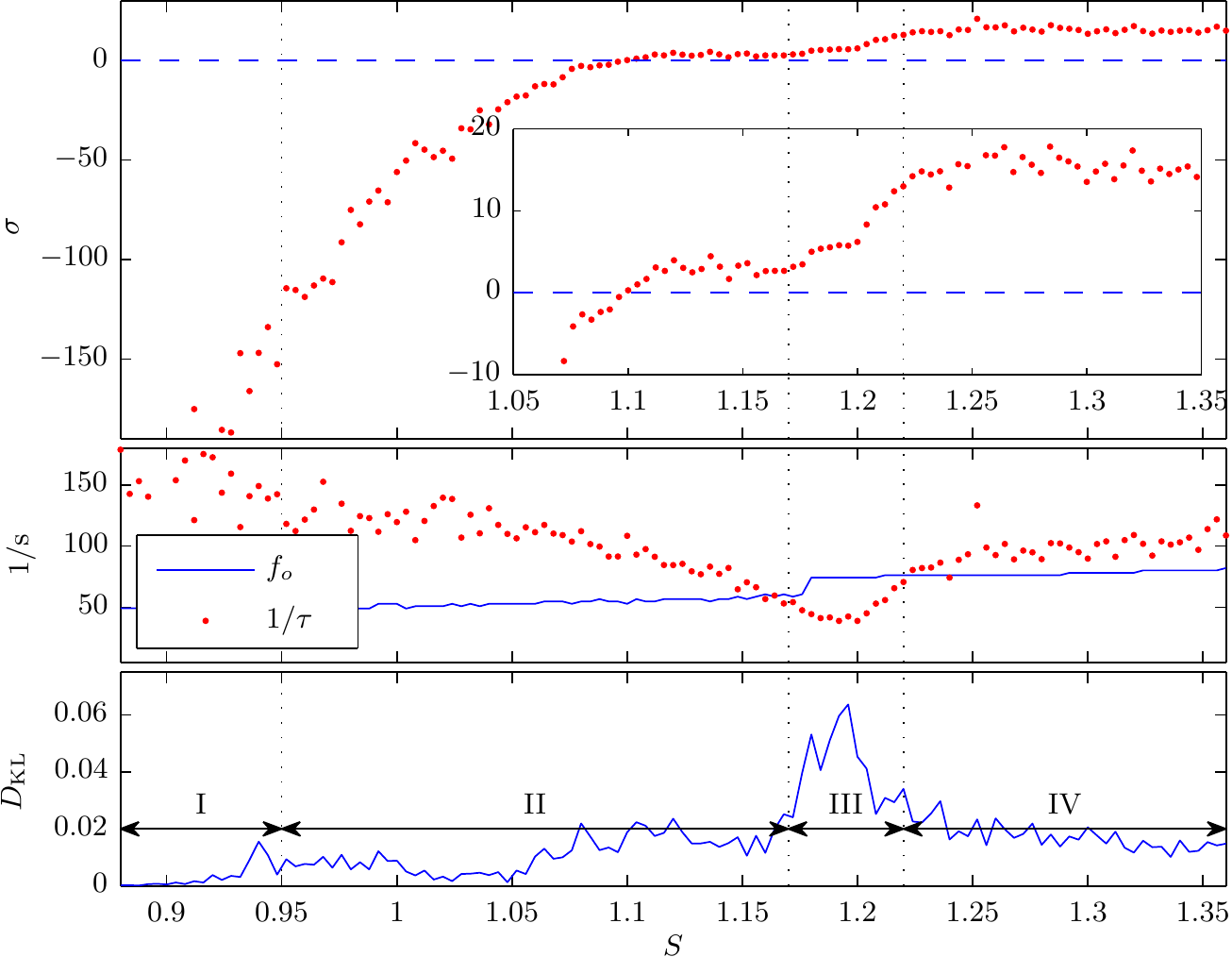}
    \caption{Estimated model parameters against the swirl number obtained from pressure measurements. The top plot gives the amplification rate and the small insert is a vertical zoom into the region around zero. The mid plot shows the oscillation frequency $f_o$ together with the noise time scale $\tau$. The lower plot shows the divergence \eqref{eqn:KLdiv} between the measured and the theoretical PDF (see fig. \ref{fig:pressure_prob_hist}). The arrows and vertical dotted lines indicate different regimes in the swirl number rage.}
    \label{fig:pressure_param_estim}
\end{figure}

This section shows the results from the application of the procedure outlined in sections \ref{sec:methods_model} to \ref{sec:methods_stat_fit} to the oscillatory mode measured from the pressure Fourier modes \eqref{eqn:fourier_decomp}. An overview of the measured PDFs is given in figure \ref{fig:pressure_prob_hist} and the estimated model parameters are presented in figure \ref{fig:pressure_param_estim}.

The amplitude PDFs presented in figure \ref{fig:pressure_prob_hist} show a continuous transition from narrow to wider distributions.
A qualitative change can be observed in regime III, where the bi-stability of the mean flow occurs.
The analytical model fits the observed PDF adequately and larger deviations are only visible in the bi-stable regime.
The gradual change of the PDF and the mean amplitude does not indicate a bifurcation point from this representation.

Figure \ref{fig:pressure_param_estim} displays the estimated amplification rate (top), the estimated noise time scale (mid) and the deviation of the PDF fit (bottom) quantified with the Kulback-Leibler divergence.
Going through the graphs from low to high swirl, the amplification rate shows first a strong increase from largely negative values up to zero. 
After the bifurcation at $S=1.1$, the estimated rates stay positive but the curve flattens and stays close to zero before there is a sudden increase in regime III.
The noise time scale is always smaller than the oscillation time scale except for regime III.
Similarly, a high divergence is observed in this regime, indicating a poor fit quality, whereas the divergence is low for all other swirl intensities. 
The overall divergence level is lowest in the stable regime and increases a little at $S=1.05$.
Overall, the presented model parameters give a clear indication of the bifurcation point and also provide a quantification of the reliability of the estimation.

The estimation results in the bi-stable regime of the flow appear not to be reliable.
This is primarily due to the stochastic switching between the two flow states that creates low-frequency perturbations of the oscillatory mode which are outside of the model's valid parameter range. 
This is distinctly indicated by the noise time scale that becomes larger than the oscillation time scale (figure \ref{fig:pressure_param_estim}, mid). 
The divergence consistently shows the failure of the model in that range.
However, the estimated amplification rate does not show large outliers in that region but rather a continuous trend between the two neighbouring regions.

The unexpected plateau of the amplification rate at swirl levels slightly above the bifurcation point may be due to the following reasons: Either the decreasing noise time scale causes an under-prediction of the estimation in confirmation with the model study (section \ref{sec:results_numeric}) or the one-dimensional dynamical model oversimplifies the underlying system dynamics at this flow regime. The latter will be addressed in the following section. 

\section{\label{sec:mean-field_model}  Modelling and system identification by the mean-field model}



\subsection{\label{sec:methods_2d_fit} Model design and calibration procedure}


In the model described in the previous section, the dynamics of the oscillatory mode are assumed to be independent of other state variable of the flow.
However, the interaction of the oscillatory mode with the mean flow might take some time such that there is a delayed saturation of the amplitude equation. 
This is further evaluated from an inspection of the mean-flow changes given by the shift-mode, which is referred by the mode coefficient $B$.
The coupled dynamics of the oscillatory mode $A$ and the shift-mode $B$ is described by the mean-field model \citep{Noack.2003, Luchtenburg.2009}, which is given as
\begin{align}
\dot{A} &= \left(\sigma + i \omega\right)A  - \beta (B-B_0) A \label{eqn:mean_field_1}\\
\dot{B} &= -\frac{1}{\tau_B}\left(B - B_0 - \gamma |A|^2\right)\label{eqn:mean_field_2}
\end{align}
in the present nomenclature.
The dynamics of the oscillatory mode $A$ is very similar to the amplitude equation \eqref{eq:landau}, but the saturation $\alpha |A|^2$ is exchanged by the feedback from the shift-mode $\beta (B-B_0) $.
The change of the shift-mode $B$ is driven by the Reynolds stresses induced by the oscillatory mode $\gamma |A|^2$.
If there is no oscillatory mode, the mean flow restores to the fixed point $B_0$ , commonly called base flow.
The rate of return to the fixed point is given by the time constant $\tau_B$.
At the limit of an infinitely short time constant, the shift-mode is slaved to the oscillatory mode as $(B-B_0)  = \gamma |A|^2$ and the amplitude equation \eqref{eq:landau} is obtained again with the Landau constant being $\alpha=\beta\gamma$.

In the present investigation, the mean-field model is adapted to capture the dynamics observed in the experimental data. 
This results in the following model, where the dimension is reduced similar to  the amplitude equation \eqref{eq:landau_pol} by changing variables to the phase-magnitude-representation and retaining only the magnitude, yielding 
\begin{align}
\dot{|A|} &= \left(\sigma- \alpha |A|^2 - \beta (B-B_0)\right) |A|+ \frac{\Gamma}{|A|} \label{eqn:2d1}\\
\dot{B} &= -\frac{1}{\tau_B}\left( B - B_0 - \gamma |A|^2 \right),\label{eqn:2d2}
\end{align}
which describes the evolution of the oscillation magnitude $|A|$ and the shift-mode $B$.

The amplitude equation in the adapted mean-field model  \eqref{eqn:2d1} covers two saturation mechanisms. 
The first is called \textit{direct saturation} and is represented by the quadratic term $\alpha |A|^2$, which is equivalent to the representation in the amplitude equation \eqref{eq:landau}. 
The second is called \textit{delayed saturation} and it is represented by the shift-mode-term $\beta (B-B_0)$ introduced in the  basic mean-field model \eqref{eqn:mean_field_1}.
The inclusion of both mechanisms is necessary to capture the dynamics observed in the data, which are discussed in the following section.

The parameters $\alpha$, $\beta$ and $\gamma$ in equation \eqref{eqn:2d1} and \eqref{eqn:2d2} are positive, which implies only negative feedback.
In other words, an increasing amplitude shifts the mean flow to a state that causes less amplitude growth and vice versa.
The fixed points of the system of equations lie on an inertial manifold called mean-field paraboloid, given by $\gamma |A|^2  = B - B_0$ \citep{Noack.2003}.
This is an attracting manifold where the mean flow state corresponds to the Reynolds stresses induced by the limit-cycle oscillations of the system.


The current mean-field model \eqref{eqn:2d1} is further adapted by the inclusion of the drift term from the amplitude equation \eqref{eq:ampl_stoch} that results from the additive noise, reading ${\Gamma}/{|A|}$.
This empirical adaption considers the effect of an 
additive noise which is not explicitly modelled for the mean-field model.
The actual addition of a stochastic forcing would need another independent noise term for the shift-mode and an extensive stochastic averaging to reduce the system to the two slow variables, the magnitude of the oscillatory mode and the shift-mode.
Furthermore, there is most probably no simple analytical form for this two-dimensional system, which would require a numerical solution of the Fokker-Planck equation \citep{Friedrich.2011}.

To overcome these difficulties, a different approach is pursued that extracts the deterministic drift directly from the statistical moments of the measured data.
Therefore, the mean-field model needs to cover only the deterministic dynamics and no stochastic forcing.
The drift is identified from a conditional average of the observed state $X$ of the stochastic process  \citep{Friedrich.2000}, reading
\begin{align}
\widetilde{f}(X_k) = \lim_{\Delta t \rightarrow 0} \frac{1}{\Delta t} \left< X(t+\Delta t) - X(t) \right>|_{X(t) = X_k} \label{eqn:CMest},
\end{align}
where $\left<\right>$ indicates the averaging operation.
The function $\widetilde{f}(X)$ is an estimate of the drift function that drives the process as given by the Langevine equation \eqref{eqn:basic_langevine}. 
In the present case, the system state is $X = [|A|,B]^T$ and the drift function $\dot{X} = f(X)$ is equations \eqref{eqn:2d1} and \eqref{eqn:2d2}.
The conditional average in equation \eqref{eqn:CMest} considers the mean drift of all trajectories that passed through a certain point in state space $X_k$, where the drift is approximated by a forward-time finite-difference.
The limit in equation \eqref{eqn:CMest} is only valid for strictly white noise, which is often not given for experimental data \citep{Lehle.2018} and generally not for turbulent perturbations.
Therefore, a fixed interval of one oscillation period was chosen to estimate the drift of the slow variables.
Neglecting the limit causes an inaccurate prediction of the drift, but relative differences and trends are still captured.

The evaluation of the conditional average in \eqref{eqn:CMest} requires the subdivision of the state space $X = [|A|,B]^T$ in bins.
For this purpose, data-based k-means clustering of the state space is pursued.
This approach has shown to be very effective for a statistical description of fluid dynamics \citep{Kaiser.2014}.
The drift, determined for each cluster centre $X_k$, is fit to the model \eqref{eqn:2d1} and \eqref{eqn:2d2} by tuning the parameters to minimise $\sum_i{(\widetilde{f}(X_k)-f(X_k))^2}$.
 To make the estimation of the noise-induced drift more robust, the parameter $\Gamma$ in \eqref{eqn:2d1} is not estimated from the drift function. 
 Instead, it is determined from the phase distortion \eqref{eq:noise_ampl_est} and scaled with a constant factor to adapt it to the drift function.
 

\subsection{\label{sec:results_2D_fit} Parameter estimation of the mean-field model from experimental data}


The parameters of the stochastic mean-field model \eqref{eqn:2d1}-\eqref{eqn:2d2} are determined from the drift coefficients estimated from the pressure measurements of the flow, according to the procedure outlined in section \ref{sec:methods_2d_fit}. 
The pressure data are processed as described in section \ref{sec:methods_identification} to obtain the magnitude of the oscillatory mode $|A|$ and the shift-mode $B$.
The data are scaled to equal variance to obtain clusters from k-means clustering with 100 centres.
Throughout the processing, the clusters have an average of \num{1500} elements and an absolute minimum of \num{24} elements.

\begin{figure}
    \centering
    \includegraphics{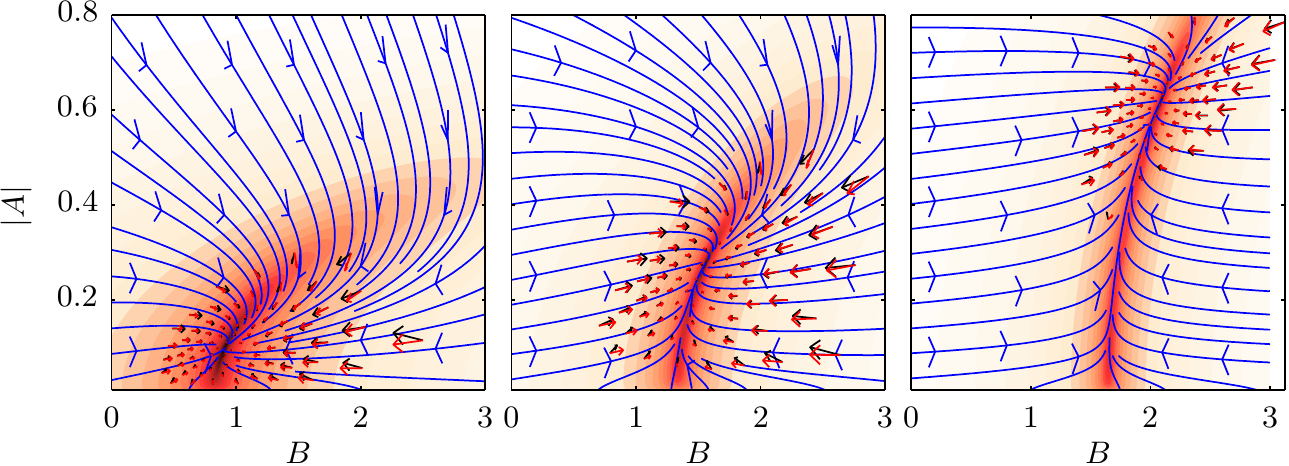}
    \caption{Fit of the stochastic mean-field model \eqref{eqn:2d1}-\eqref{eqn:2d2} to the estimated drift coefficients \eqref{eqn:CMest}. Black arrows indicate the estimated coefficients and red arrows show the model coefficient at the same point. The streamlines and the contour in the background show the global picture of the model drift direction and magnitude, respectively (drift: $f(X)=\dot{X}$ with $X = [|A|,B]^T$ ). Fields for $S=[0.96,1.12,1.28]$ are depicted from left to right.}
    \label{fig:state_space_2D}
\end{figure}

The drift coefficients estimated for each cluster centre and the fitted mean-field model are displayed in figure \ref{fig:state_space_2D}.
The model shows some deviations from the measurement data, especially at states far away from the fixed point. 
However, the general trend is very accurately captured.
The model shows the attraction of the flow to the fixed point and also the mean-field paraboloid that appears as a bent region with low drift magnitudes.
The three displayed cases correspond to the stable and the two unstable regimes, respectively.
Despite their different dynamic states, the models show very similar dynamics in state space.
The main change in the drift field is due to the increase of the limit-cycle amplitude and the corresponding movement of the fixed point in state space to larger amplitudes $|A|$.
A specific characteristic of the present model is visible from the horizontal approach towards the fixed point.
Especially for the intermediate case (figure \ref{fig:state_space_2D} mid), there is an upward trend on the left side and a downward trend on the right side.
This indicates the dependence of the oscillatory mode dynamics on the shift-mode, otherwise, there must be horizontal drift lines at the limit-cycle amplitude.
The general agreement of the observed and modelled dynamics is also a validation of the empirical mean-field model.
The model is designed such that it covers the shown dynamics by combining elements from the stochastic amplitude equation and a basic mean-field model.
Therefore, it indicates the dynamics which are relevant for the observed flow.

\begin{figure}
    \centering
    \includegraphics{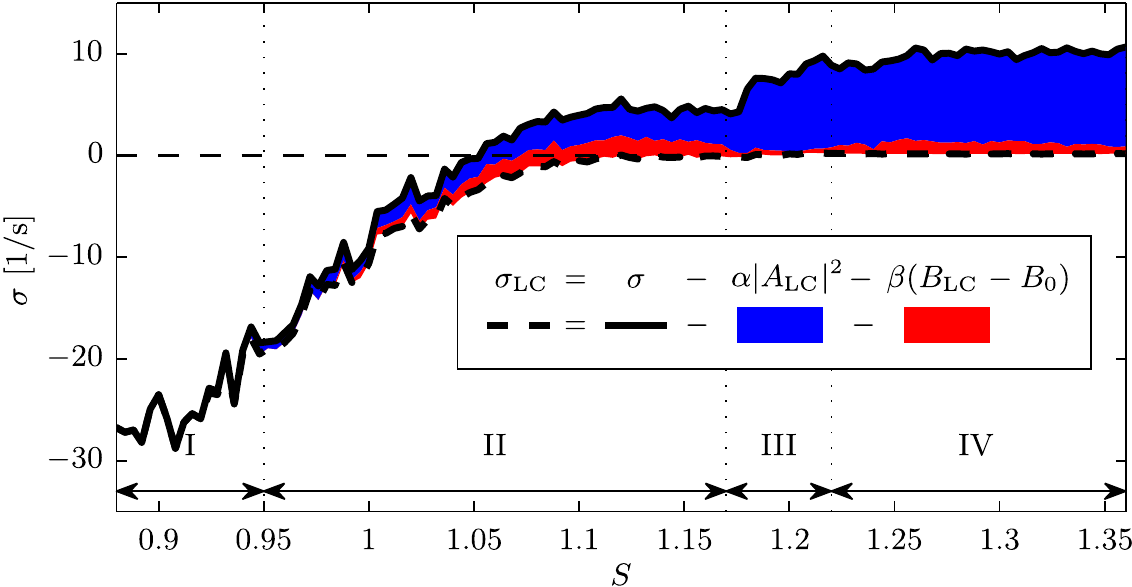}
    \caption{Estimated model parameters from pressure measurements using the mean-field model. The coloured areas indicate different contributions to the saturation of the amplification rate in \eqref{eqn:sigma_LC}. The arrows and vertical dotted lines indicate different regimes in the swirl number range.}
    \label{fig:sigma_2D}
\end{figure}

Figure \ref{fig:sigma_2D} shows the estimated amplification rate obtained from the calibration of the mean-field model.
The graph gives the linear amplification rate similar to figure \ref{fig:pressure_prob_hist} and further details which mechanism in the model leads to the saturation at the limit-cycle.
The coloured areas indicate different contributions to the saturation of the amplification rate at the limit-cycle as given in equation \eqref{eqn:2d1}. Accordingly, the effective amplification rate at the limit-cycle is given by 
\begin{align}
    \sigma_{\mathrm{LC}}=\sigma- \alpha |A_\mathrm{LC}|^2  - \beta (B_\mathrm{LC}-B_0),\label{eqn:sigma_LC}
\end{align}
where the subscript LC refers to a specific amplitude at the limit-cycle.
The two contributions that lead to the reduction of the initial linear amplification rate $\sigma$ to the amplification rate at the limit-cycle $\sigma_{\mathrm{LC}}$ are shown.
The difference due to the direct saturation $\alpha |A_\mathrm{LC}|^2$ and the delayed saturation $\beta (B_\mathrm{LC}-B_0)$ are indicated in the graph.
In contrast to the estimates from the amplitude equation (figure \ref{fig:pressure_param_estim}), which showed a bifurcation at $S=1.1$, the mean-field model shows a bifurcation of the flow at $S=1.05$.
The graph further shows that the delayed saturation is much more relevant in regime II.
In regime IV, the main contribution comes from the direct saturation term in the model.
For the larger swirl numbers, the two saturation mechanism always add up to neutral stability at the limit-cycle ($\sigma_\mathrm{LC}\approx0$).
Furthermore, the magnitude of the linear amplification rate $\sigma$ in regime II goes clearly above the zero line in contrast to the previous estimation from the amplitude equation (figure \ref{fig:pressure_param_estim}).
This is more consistent with the continuous increase of the limit-cycle amplitude observed in figure \ref{fig:pressure_prob_hist}.

The agreement of the observed and modelled drift coefficients, as well as the consistent description of the flow physics, suggest a more reliable description of the bifurcation by the mean-field model than the amplitude equation. Especially at the bifurcation point, the secondary dynamics from the shift-mode contribute significantly to the dynamics of the oscillatory mode, which causes the deviation of the more simple approach based on the  amplitude equation.




\subsection{\label{sec:results_2D_discussion} Insights from the description of the flow by the mean-field model}


The parameters estimated from the 2D mean-field model \eqref{eqn:2d1}-\eqref{eqn:2d2} allow an accurate description of the dominant flow dynamics in consistency with the 1D amplitude equation \eqref{eq:ampl_stoch}. 
The 2D model, however, differs in two main points: (i) The identified bifurcation point and (ii) the growth rates for swirl numbers closely beyond the bifurcation point. The 2D model reveals that the interaction of the shift-mode and the oscillatory mode in this regime is of large importance, which is not covered by the 1D model.
Hence, neglecting the shift-mode and describing the system solely based on the amplitude makes it a non-Markovian process, which can not be covered by the proposed Langevine equation.
The unresolved dynamics between the shift-mode and oscillatory mode cause the additive noise to be correlated with the dynamics, which violates the basic assumptions.

\begin{figure}
    \centering
            \includegraphics[width=0.95\textwidth]{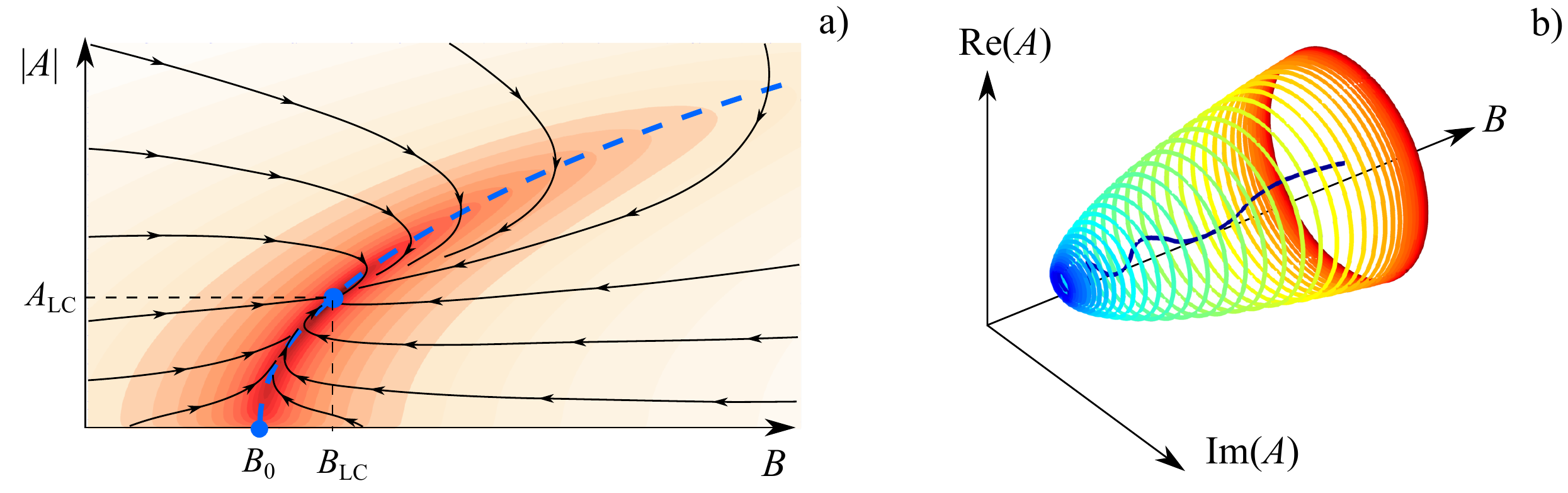}
    \caption{a) Reduced schematic of one of the plots from figure \ref{fig:state_space_2D}, where the mean-field paraboloid is indicated as dashed line. The base flow is indicated by $B_0$ and the mean flow (limit-cycle) by $B_\mathrm{LC}$. b) Simulated time series of the mean-field model for $S=1.12$, where the line is coloured by the simulation time (from blue to red). The time series is starting at $B=B_\mathrm{LC}$ and $A=0$, the unperturbed mean-flow.}
    \label{fig:mean_field_model}
\end{figure}

The results from the calibration of the mean-field model provide a deeper understanding of the dynamics governing the swirling flow. 
The corresponding properties of the flow model are summarised in figure \ref{fig:mean_field_model} a.
Two distinct flow states can be derived from the model: (i) The base flow, which constitutes the quasi-stationary state of the flow before the onset of the instability; and (ii) the mean flow, which is the mean velocity field corresponding to the instability oscillating at the limit-cycle.
The base flow is a state that is rarely reached in unstable conditions due to the unavoidable external perturbations from turbulence and instabilities. It can only be obtained artificially through active flow control or from transient investigations of the flow.
The unique feature of the current approach is that one can make statements about the stationary base flow state, although one considers only fluctuations around the mean flow state.

More generally, the better understanding of the interaction between hydrodynamic instabilities and the mean flow helps to link the experimental observations and numerical results from mean-flow stability analysis beyond the comparison of mode shapes.
The quantitative assessment of amplification rates and mean-flow corrections from measurements makes the results from both approaches directly comparable.
Moreover, the distinction between deterministic and stochastic parts and their contribution to the flow dynamics enriches the picture of hydrodynamic instabilities in turbulent flows and helps to interpret the amplification rates obtained from mean-flow stability analysis.
The effective limit-cycle amplification rate (see figure \ref{fig:sigma_2D}) are directly comparable to the amplification rate of a corresponding mean-flow stability analysis.

The simulation of a time series shown in figure \ref{fig:mean_field_model} b, starting from the mean flow with no oscillations, further illustrates the essence of the mean-field model. 
Before the oscillations grow significantly, the flow rapidly shifts towards the base flow and then grows in amplitude along the mean-field paraboloid. 
This behaviour is consistently observed for simulations of the cylinder wake \citep{Noack.2003,Brunton.2016}, where the mean-field paraboloid is also identified as an inertial/slow manifold.
The drift in state space (figure \ref{fig:mean_field_model} a) clearly shows that the mean-field paraboloid constitutes an attracting, slow manifold for the swirling jet.
This knowledge of the mean-field model helps to understand the transient and intermittent dynamics of the swirling jet.

\section{\label{sec:conclusions}Conclusions}
In this work, a method was developed to estimate the properties of a global hydrodynamic instability from measurement data of turbulent flows. 
The approach makes use of the stochastic perturbations that are present in the flow due to background turbulence. 
Background turbulence pushes the flow away from stable fixed-points or limit-cycles and, thus, forces the dynamics into other parts of the state space. From the deterministic return to the fixed-point or limit-cycle, the dynamic properties of the flow can be extracted.

The dynamical system is modelled by a stochastic amplitude equation describing the oscillatory dynamics of the instability (1D model) and, in a second approach, by a stochastic mean-field model that captures additionally the interaction between the instability and the mean-flow corrections (2D model). The stochastic perturbations are incorporated as additive forcing.

To capture the spectral properties of the turbulent perturbations, coloured noise was used for the stochastic forcing in the  dynamical system.
The validity of the derived amplitude equation for coloured noise was investigated by a numerical study. 
It is shown that the approach is feasible as long as the noise time-scale is smaller than the oscillation period of the instability. 

The methodology was applied to experimental data of a turbulent swirling jet undergoing vortex breakdown. This flow is dominated by a helical global mode commonly termed the precessing vortex core. Thereby, the swirl number is the major control parameter that governs the supercritical Hopf bifurcation of the global mode. 
PIV measurements were conducted to ensure that this mode is the most dominant coherent structure in the flow for the investigated swirl number range. 
For the system identification, pressure measurements around the nozzle lip were used providing longer time series than PIV.

The application of the 1D model showed very good capabilities to fit the observed dynamics of the flow.
The bifurcation point of the global mode was identified and the good agreement between measured and estimated statistics showed that the model captures the relevant dynamics.
The approach also identifies regions of a potential mismatch between the modelled and observed dynamics.
The occurrence of an intermediate bi-stable switching between two flow states was correctly identified as a regime that is not accurately  captured by the model.
Moreover, the 1D model predicts a plateauing of the  growth rate  of the instability shortly beyond the bifurcation point, which contradicts the continuous increase of the limit-cycle amplitude.
This discrepancy is addressed in the 2D mean-field model.
Nevertheless, the 1D approach is fairly robust with regard to increased noise magnitude and noise time-scales.

The description of the flow from a stochastic mean-field model was introduced as an alternative estimate of the flow properties.
The calibration of the 2D model provides a coarse estimate since no reduced analytical solution was derived.
In contrast to the analytical approach pursued for the 1D amplitude equation, the drift coefficients were determined from statistical moments of the measurement data.
Other than the development of the equation for the amplitude PDF in the first approach, the estimate of the drift coefficients need no prior knowledge of the flow model. 
The 2D model was constructed a posterior to correspond to the observed dynamics. 
This will serve as a point of departure for future developments of a stochastic mean-field model that also incorporates accurate stochastic forcing and consequent development of 2D analytical models from stochastic averaging.

The main difference in the description of the flow by the 2D instead of 1D model, is a bifurcation of the flow at slightly lower swirl numbers and an increased growth rate right after the bifurcation.
These differences most probably arise from not accounting for the interaction between the oscillatory mode and the slow mean-flow corrections by the shift-mode.
In the 1D model, these are lumped into the stochastic forcing which violates the assumption of purely additive forcing.
The consideration of the of the mean-flow corrections by the 2D model clarifies this transient dynamic of the flow.
This gives a more detailed picture of the flow dynamics and allows an estimation of the unperturbed base-flow state.

The work demonstrates that the observation of limit-cycle oscillations is not sufficient to determine the flow state as the influence of the stochastic turbulent forcing is significant and masks the actual bifurcation point. However, the proposed separation of deterministic and stochastic contributions in the dynamical model allows identifying the flow sate solely based on stationary measurement data.
The inclusion of the shift-mode gives further capabilities to handle flows with a pronounced mean-flow correction.
The methodology is expected to apply to a wide range of turbulent flows subjected to global flow instabilities.

\begin{appendix}

\section{\label{sec:swirl_number}Swirl number determination}

\begin{figure}
    \centering
    \includegraphics{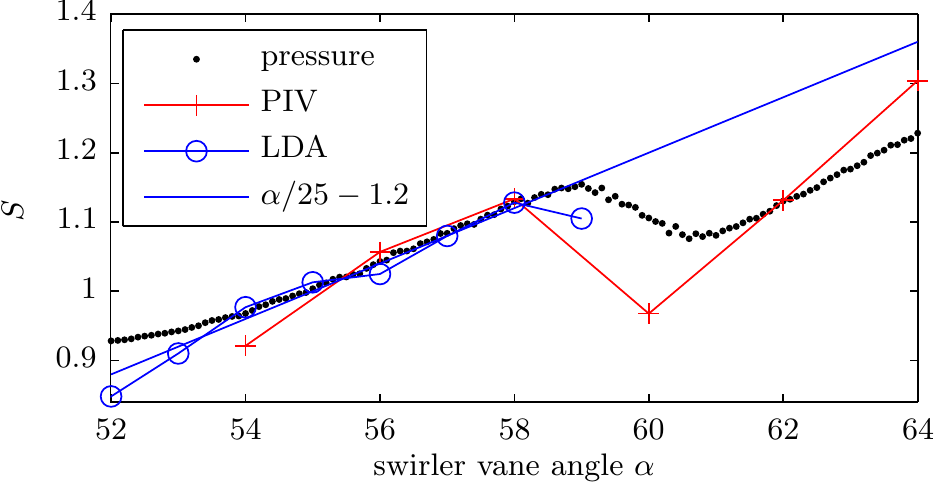}
    \caption{Different swirl number definitions against the swirler vane angle of the experimental apparatus. The integral swirl numbers are obtained from PIV and Laser Doppler Anemometer (LDA) measurements. Approximations of the integral swirl number based on the pressure measurements and the swirler angle are given as well.}
    \label{fig:swirl_number}
\end{figure}

The swirl number for the present investigations is obtained from PIV and LVD measurements as presented in figure \ref{fig:swirl_number}.
The integral swirl number is computed as in \cite{Oberleithner.2012} from the ratio of axial flux of azimuthal to axial momentum 
\begin{align}
    \mathrm{S} = \frac{2 \dot{G_{\theta}}}{\mathrm{D} \dot{G_{x}}} = \frac{2\pi  \int\limits_{0}^{\infty}\rho  \overline{v}_{x} \overline{v}_{\theta}  r^{2} \mathrm{d}r}{\mathrm{D}  \pi \int\limits_{0}^{\infty}\rho \left(\overline{v_{x}^{2}}- \frac{\overline{v_{\theta}^{2}}}{2}\right)  r \mathrm{d}r}.
\end{align}
The general perception is that the integral swirl number is very sensitive to the axial position utilised for the calculations, although it should be constant along the jet axis.
This is due to difficulties of an accurate representation of the pressure-related momentum transport in a turbulent flow.
Especially the presence of stagnant or reversed flow at the axial position of evaluation causes complications.

Since the current investigations span a large range of swirl numbers, there are cases where the recirculation region reaches up to the nozzle and different breakdown shapes occur.
The graphs in figure \ref{fig:swirl_number} reflect these properties, where different  measures of the swirl number are plotted against the swirl generator vane angel.
Due to the transition between the two mean flow states, there is a jump in the representation of swirl number against the swirler vane angle.
As the integral swirl number is not unique for the investigated range, a swirl number based on the swirl angle is calibrated to the integral swirl number resulting in $\mathrm{S}_{\alpha} = \alpha/25^\circ - 1.2$.
This geometry based swirl number is used throughout the current investigation.

The black dots in figure \ref{fig:swirl_number} represent a swirl measure based on the time mean $m=0$ pressure Fourier mode $\left<\widehat{p}_{0}\right>$ as $\mathrm{S}_{\mathrm{p}} = \left<\widehat{p}_{0}\right>/28\mathrm{Pa} + 0.9$.
It becomes clear that that the pressure is a good indicator for the mean flow state, which justifies the description of the shift-mode being proportional to the mean pressure.
The relation is only violated in the bi-stable region where the model predictions fail anyway.


\section{\label{sec:PIV_pressure_relation} Relation between PIV and pressure measurements}

\begin{figure}
    \centering
    \includegraphics{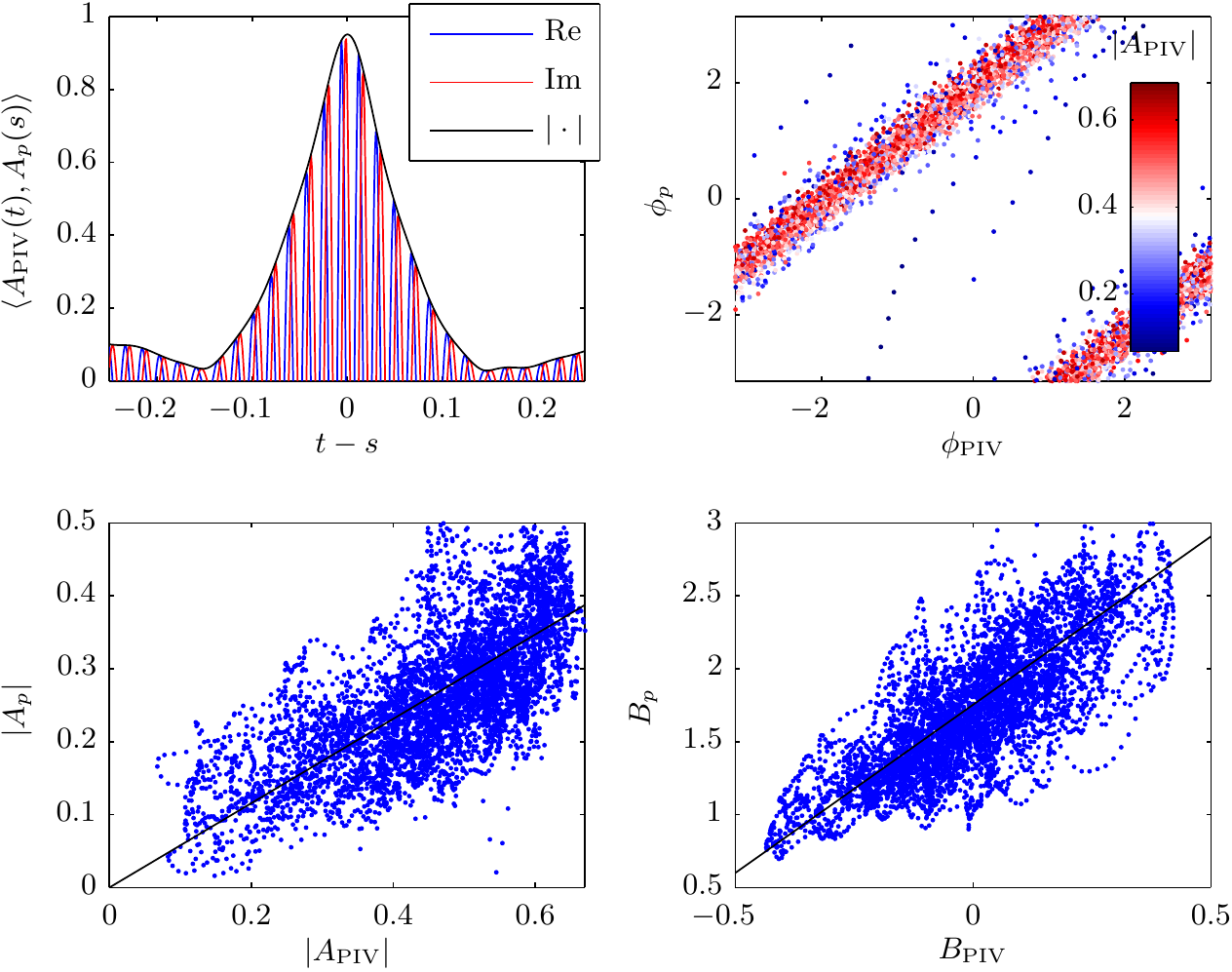}
    \caption{Comparison of the helical mode coefficient obtained from simultaneous PIV and pressure measurements as indicated by the subscripts PIV and $p$, respectively. (top left) The correlation coefficient between the complex coefficient from both measurements. (top right) Relation of the phases (argument of the complex amplitude), the colour indicates $|A_{\mathrm{PIV}}|$. (bottom left) Relation of the helical magnitudes and (bottom right) relation of the shift-modes, where the linear trend is indicated by black lines.}
    \label{fig:PIV_pressure_relation}
\end{figure}

In this work, the dynamics of the helical mode and the shift-mode are quantified based on velocity and pressure data. To confirm the correlation between these two quantities, the data from simultaneous PIV and pressure measurements of the flow are analysed and presented in figure \ref{fig:PIV_pressure_relation}.
The presented data are from a \SI{5.5}{\second} measurement series, where synchronised measurements are recorded at a swirl number of $S=1.12$.
The plot compares the oscillatory mode $A$ and the shift-mode $B$ obtained (i) from the pressure measurements as Fourier decomposition of the pressure signals  as $A_p = \widehat{p}_1$ and $B_p = \widehat{p}_0$ and (ii) from the SPOD coefficients of the PIV measurements as $A_\mathrm{PIV} = a_1 + i a_2$ and $B_\mathrm{PIV} = a_3$.
The phase and magnitude of the oscillatory mode are obtained from the polar representation of the complex coefficient, reading $A=|A|\exp(i\phi)$.

The cross-correlation (figure \ref{fig:PIV_pressure_relation} top left) indicate a good overall agreement of the helical mode obtained from botch approaches, where the maximum correlation coefficient is \num{0.95}.
Since both measurements are taken at different locations, there is a \SI{5}{\milli\second} shift in time between PIV and pressure data, corresponding to a quarter oscillation period of the flow oscillation. 
The maximum of the correlation is used to align both measurement series in time.

The direct comparison of the phase (figure \ref{fig:PIV_pressure_relation} top right) and the magnitude (figure \ref{fig:PIV_pressure_relation} bottom left) allows a more detailed investigation of this relation.
The phases show limited jitter with a standard deviation of less than 5\% of a period and a very good agreement. 
Relatively larger deviations are mainly observed at times with low oscillation magnitude.

The magnitude comparison (figure \ref{fig:PIV_pressure_relation} bottom left) indicates larger deviations between the two measures, especially the pressure shows stronger fluctuation around the limit-cycle. 
This is because the SPOD coefficient of the PIV data represents the average magnitude over a large measurement domain, whereas the pressure measurement only senses the local dynamics near the nozzle lip.
Thus, any local perturbations of the helical mode is spatially averaged in the SPOD coefficient but directly visible in the pressure signal.
Moreover, the large difference in the number of spatial measurement points causes a larger pickup of measurement noise in the pressure estimation of the mode coefficient.

Figure  \ref{fig:PIV_pressure_relation} (bottom right) shows the corresponding comparison of the shift-mode from both measurements.
Similar to the magnitudes, the pressure estimation shows larger fluctuations, but the average dynamics agree very well.

\end{appendix}

\section*{Acknowledgment}\label{acknowledgment}
The authors kindly acknowledge the funding from the German Research Foundation under DFG Project PA 920/30-1 and DFG Project PA 920/37-1. 

\section*{Declaration of interests}
The authors report no conflict of interest. 

\bibliographystyle{jfm}
\bibliography{references}

\end{document}